\documentclass[11pt,a4paper]{article}
\usepackage{jheppub}
\pdfoutput=1

\usepackage{amsmath,bbm,latexsym,url,color}
\usepackage{graphicx}
\usepackage{color}
\usepackage[normalem]{ulem}
\usepackage{bbm}

\definecolor{dgreen}{rgb}{0.00,0.50,0.00}

\numberwithin{equation}{section}

\newcommand{\nc}{\newcommand}
\nc{\non}{\nonumber}
\nc{\hc}{\hbox {H.c.}} 
\nc{\noi}{\noindent}
\nc{\barx}{\bar{x}}
\nc{\pb}{\;\hbox {pb}}
\nc{\fb}{\;\hbox {fb}}
\nc{\mev}{\;\hbox {MeV}}
\nc{\gev}{\;\hbox {GeV}}
\nc{\tev}{\;\hbox {TeV}}
\nc{\lsp}{\;\;\;\;\;}
\nc{\Lsp}{\;\;\;\;\;\;\;\;\;\;}  
\nc{\LLsp}{\lspace \lspace}
\nc{\lra}{\longrightarrow}
\nc{\llra}{\longleftrightarrow}

\nc{\beq}{\begin{equation}}  \nc{\eeq}{\end{equation}}
\nc{\bea}{\begin{eqnarray}}  \nc{\eea}{\end{eqnarray}}
\nc{\baa}{\begin{array}}     \nc{\eaa}{\end{array}}
\nc{\bit}{\begin{itemize}}   \nc{\eit}{\end{itemize}}
\nc{\ben}{\begin{enumerate}} \nc{\een}{\end{enumerate}}
\nc{\bce}{\begin{center}}    \nc{\ece}{\end{center}}
\nc{\bpm}{\begin{pmatrix}}   \nc{\epm}{\end{pmatrix}}
\nc{\bvt}{\begin{verbatim}}  \nc{\evt}{\end{verbatim}}
\def\lsim{\mathrel{\raise.3ex\hbox{$<$\kern-.75em\lower1ex\hbox{$\sim$}}}}
\def\gsim{\mathrel{\raise.3ex\hbox{$>$\kern-.75em\lower1ex\hbox{$\sim$}}}}
\nc{\hp}{\hat\phi}
\nc{\tp}{\tilde\phi}
\nc{\vp}{\vec\varphi}
\nc{\mvp}{m_\varphi}
\nc{\mh}{m_h}
\nc{\lvp}{\lambda_\varphi}
\nc{\lh}{\lambda_H}
\nc{\lx}{\lambda_x}
\nc{\dm}{{\Omega_{\mathrm{DM}}^{\mathrm{(exp)}}}}
\nc{\Dh}{\Delta_h}

\nc{\for}{\;\;{\rm for}\;\;}
\nc{\then}{\;\;{\rm then}\;\;}
\nc{\aaa}{\;\;{\rm and}\;\;}

\nc{\p}{\partial}

\def\sv{\left\langle\sigma v\right\rangle}

\definecolor{purple}{rgb}{.6,0.4,1}
\definecolor{rose}{rgb}{0.93,0.8,0.8}

\begin{document}

\title{Multi-Scalar-Singlet Extension of the Standard Model\\
- the Case for Dark Matter and  an Invisible Higgs Boson - }


\preprint{IFT-11-9,  UCRHEP-T514}
1112.2582

\author[a]{A. Drozd}
\emailAdd{Aleksandra.Drozd@fuw.edu.pl}
\author{B. Grzadkowski}
\emailAdd{Bohdan.Grzadkowski@fuw.edu.pl}
\affiliation[a]{Institute of Theoretical Physics, University of Warsaw, 
Ho\.za 69, PL-00-681 Warsaw, Poland}
\author{Jos\'e Wudka}
\emailAdd{Jose.Wudka@ucr.edu}
\affiliation[b]{Department of Physics, University of California,
Riverside CA 92521-0413, USA}

\abstract{
We consider a simple extension of the Standard Model by the addition of $N$ real scalar
gauge singlets~$\vp$ that are candidates for Dark Matter. By collecting theoretical and 
experimental constraints we determine the space of allowed parameters of the model.
The possibility of ameliorating the little hierarchy problem within the multisinglet model is discussed.
The Spergel-Steinhardt solution of the Dark Matter density
cusp problem is revisited. 
It is shown that fitting the recent CRESST-II data for Dark Matter nucleus scattering implies
that the standard Higgs boson decays predominantly into pairs of Dark Matter scalars.
It that case discovery of the Higgs boson at LHC and Tevatron is impossible.
The most likely mass of the dark scalars is in the range $15\gev \lsim \mvp \lsim 50\gev$ with
$BR(h\to \vp\vp)$ up to $96\%$. 
}

\keywords{dark matter, gauge singlet scalars, Higgs boson decay} 
\arxivnumber{1112.2582} 

\maketitle

\section{Introduction}
\label{intro}

The evidence for dark matter (DM) has now become almost overwhelming~\cite{Jarosik:2010iu}, yet
the standard model (SM) does not contain a viable DM candidate, which necessitates a modification of this model. 
The simplest possible such extension consists of introducing a real scalar field which is a gauge singlet.
Many authors considered this singlet extension of the SM; the earliest publication we are aware of
was by Veltman and Yndurain in \cite{Veltman:1989vw} (though their
motivation was different). The DM issue was first addressed  in this context in
\cite{Silveira:1985rk} 
and  \cite{McDonald:1993ex} and then followed by other
authors \cite{Burgess:2000yq}-\cite{Pospelov:2011yp}.

In this paper we will consider a generalization of this model
by extending the DM sector to an unbroken (global) $O(N)$ model. We will
derive relevant theoretical and experimental constraints on this model 
and determine range of allowed parameters. Since we 
demand the new singlets do not acquire a vacuum expectation value,
the effects of the dark sector on precision observables is
much suppressed, appearing only at the two (or higher) loop level. The only experimental constraints are 
then obtained from the requirement that the scalars are adequate candidates for DM. The theoretical 
constraints on the model are derived from vacuum stability, perturbativity, and triviality.
In addition we will also determine the fine tuning conditions that ameliorate
 the little hierarchy problem by seeking model parameters that 
 tame quadratically divergent contributions to the Higgs boson mass~\cite{Grzadkowski:2009mj},
\cite{Kundu:1994bs} following strategies adopted in \cite{Kolda:2000wi}.

The model discussed below is also interesting in light of the DM signals
recently announced by the CRESST-II collaboration~\cite{Angloher:2011uu}. 
We will see that, not only the most likely regions found by CRESST-II are consistent with the model, but in a region of parameter space
where the SM Higgs decays invisibly into DM pairs. Such a Higgs boson 
would not be detected at either the 
Tevatron or LHC. 

The paper is organized as
follows. In Sec.~\ref{Sec:model} we introduce the model and formulate strategy adopted in
this paper for its study. In Sec.~\ref{Sec:stab-uni} we review the vacuum stability and unitarity bounds that restrict 
the parameter space of the model, while in Sec.~\ref{Sec:triv} the issue of triviality is addressed.
Sec.~\ref{Sec:fin-tun} is devoted to a discussion of one- and leading two-loop corrections
to the Higgs boson mass in the presence of extra singlets.
In Sec.~\ref{Sec:DM} the present Dark Matter abundance is investigated for the Cold Dark Matter (CDM) and 
Feebly Interacting Dark Matter (FIDM) scenarios.
Sec.~\ref{Sec:dir-det} shows results of constraining the model by results of direct searches of DM. 
In Sec.~\ref{Sec:self-int} we revisit the Spergel-Steinhardt hypothesis of self-interacting DM.
Sec.~\ref{Sec:sum} contains summary and conclusions.

\section{The Model}
\label{Sec:model}

We  consider the SM of electroweak interactions extended by the addition of $N$ scalars $\vp$ that are singlets under the SM gauge group $SU(3)\times SU(2)\times U(1)$, and transforming according to the fundamental representation of $O(N)$, under which all SM fields are singlets; for simplicity we assume that $O(N)$ is an exact symmetry of the model~\footnote{This condition could be relaxed without significantly altering the conclusions; the only consequence would be the presence of many more parameters of the model.}. With the intention of providing a Dark Matter (DM) candidate we impose an additional $Z_2$ symmetry, which is not spontaneously broken, and under which $\vp$ is odd: $\vp\to-\vp$ while all other fields are even~\footnote{For $N>1$ this $Z_2$ symmetry is a consequence of $O(N)$ invariance.}.
The most general, symmetric and renormalizable potential reads:
\begin{eqnarray}
V(H, \vec{\varphi}) = 
- \mu_{H}^2 H^{\dagger} H 
+ \lambda_{H} (H^{\dagger} H)^2 
+ \frac{1}{2}  \mu_{\varphi}^2 \vec{\varphi}^2
+ \frac{1}{4!} \lambda_{\varphi} \left(\vec{\varphi}^2 \right)^2
 +  \lambda_{x} H^{\dagger} H \vec{\varphi}^2 \,,
 \label{pot}
\end{eqnarray}
where $H$ is the SM $SU(2)$  Higgs isodoublet. 
The Lagrangian density for the scalar sector is then given by:
\begin{eqnarray}
L_{scalar} = \frac{1}{2} \partial_{\mu} \vec{\varphi} \partial^{\mu} \vec{\varphi} + D_{\mu}H^{\dagger} D^{\mu} H 
- V(H, \vec{\varphi})\,.
\end{eqnarray}
As usually the minimum of the potential breaks spontaneously electroweak symmetry via non-zero vacuum expectation value of the Higgs doublet $\left<H\right> = (0, v/\sqrt{2})$, $v = 246 \gev$. Since we require the $O(N)$ symmetry  ($Z_2$ for $N=1$) to remain unbroken, we assume that $\mu_\varphi^2 > 0$, so $\left<\vp\right> = 0$. Note that $\left<\vp\right> = 0$ implies no mass-mixing between 
$\vp$ and $H$, therefore the existing collider limits on the Higgs properties are not modified.
After the symmetry breaking the physical scalars have masses 
$\mh^2 = - \mu_{H}^2 + 3 \lambda_{H} v^2 =2 \mu_{H}^2$ 
and
$m_{\varphi}^2 = \mu_{\varphi}^2 + \lambda_{x} v^2 $. Note that all components of $\vp$ have the same mass as 
the consequence of $O(N)$, this degeneracy can be removed by adding a generic mass term $ (\mu_\varphi)_{ij} \varphi_i \varphi_j $
that only breaks $O(N)$ softly.

The model then contains five unknown parameters: $\mh, \mvp, \lx, \lvp, N$. Our goal is to constrain the parameters taking into account available restrictions: theoretical (vacuum stability, unitarity/perturbativity, triviality of the scalar sector, Higgs mass correction fine-tuning) and experimental (DM relic abundance, direct detection experiments).

\section{Stability and perturbative unitarity}
\label{Sec:stab-uni}

In order to stabilize vacuum we will assume that the scalar potential (\ref{pot}) is (at tree level) bounded from below. This condition implies
\beq
\lambda_H\,, ~\lambda_{\varphi} > 0  \,; \quad \lambda_x > - \sqrt{\frac{\lambda_{\varphi}\lambda_{H}}{6}}  =  -\frac{\mh}{2v} \sqrt{ \frac{\lambda_{\varphi}}{3}} \,.
\label{stab_con}
\eeq
 
Tree-level unitarity constraints emerge from the SM condition for $V_LV_L$ scattering~\cite{Lee:1977eg} and from the requirement that all possible scalar-scalar scattering amplitudes are consistent with unitarity of the $S$ matrix \cite{Cynolter:2004cq} 
\beq
\mh^2< \frac{8 \pi}{3} v^2, \lsp \lambda_{\varphi} < 8 \pi \aaa |\lambda_x| < 4 \pi\,.
\label{unit_con}
\eeq

Finally, the condition that the global $O(N)$ symmetry remains unbroken 
requires $\mu_{\varphi}^2 > 0$ which leads to the very useful inequality: 
\beq
m_{\varphi}^2 > \lambda_{x}v^2 \,;
\label{mphi_bound}
\eeq
a consequence is that  light scalars ($\mvp \ll v$) must couple very weakly to the SM
($\lx \ll 1 $).

\section{Triviality} 
\label{Sec:triv}

The 'triviality bound' is a constraint on the SM generated by the requirement that 
under the renormalization group evolution the quartic coupling constant $\lh$ remains finite 
up to the UV cut-off scale $\Lambda$ of the model; in other words, 
that the Higgs-boson Landau pole should be above $\Lambda$. This requirement implies conditions on the initial values of various running parameters of the model, and in particular on 
$\lh(\mu= m_W)$; this, in turn, leads to an upper limit on the Higgs boson mass as a function of $\Lambda$. 

In order to determine the location of the Landau pole as a function of the Higgs mass
one has to solve the renormalization group evolution (RGE) equations for all of the running parameters of the model. We will be interested only in cutoff scales below $ 50 \tev$, in which case the RGE
of the gauge and top-quark Yukawa couplings can be safely neglected. Thus we
only need to consider the evolution of $\lambda_H$, $\lambda_\varphi$ and $\lambda_x$ as determined by
\begin{eqnarray} 
16 \pi^2 \mu \frac{d \lh}{d \mu} &=& 
\frac{3}{8}  g_{1}^4 + \frac{9}{8} g_{2}^4 + \frac{3}{4} g_{1}^2 g_{2}^2 
- 6 y_{t}^4 + 24 \lh^2 + 12 y_{t}^2 \lh -3 g_{1}^2 \lh - 9 g_{2}^2 \lh 
\nonumber\\ & & 
+ 2 N\lx^2 \label{diff_set1}\\
16 \pi^2 \mu \frac{d \lx}{d \mu} &=& 
\lx\left( 12 \lh + \lvp + 8 \lx + 6 y_{t}^2 - \frac{3}{2} g_{1}^2 - \frac{9}{2} g_{2}^2\right)  
\label{diff_set2}
\\ 
16 \pi^2 \mu \frac{d \lvp}{d \mu} &=& 48 \lx ^2+ \frac{1}{3}(8 + N)\lvp^2
\label{diff_set3}
\end{eqnarray}  
where $g_1$, $g_2$, $g_3$ are the gauge coupling for $U(1),SU(2), SU(3)$, respectively, and $y_t$ is the top quark Yukawa coupling Unique solutions could be obtained once the initial conditions
\begin{eqnarray} 
\lambda_H(\mu = m_W)&=& \lambda_{H\, 0} \label{initial1}\\
\lambda_{x}(\mu = m_W)&=& \lambda _{x \, 0} \label{initial2}\\
\lvp(\mu = m_W)&=& \lambda _{\varphi \, 0} \label{initial3}
\end{eqnarray}
are specified.

For a given $\Lambda$ we shall require that there is no pole in the evolution of scalar quartic coupling constants at 
energies below $\Lambda$. The region in the $(\Lambda, m_{h}= v \sqrt{2\lambda_{H\,0}})$ plane 
allowed by this constraint, depends on the initial parameters $\lambda_{x \, 0}$, $\lambda _{\varphi \, 0}$, and the number of scalars $N$. For each $ \Lambda $ the maximum allowed value of $ \mh $
constitutes the triviality bound on the Higgs mass. One might worry that each quartic coupling constant 
will have a different pole location, this, however, does not occur when $\lambda_{x \, 0} \neq 0$ 
(as we assume), for then all the RGE equations
are coupled and $ \lambda_{H,\varphi,x} $ diverge at the same value of $ \mu $.

\begin{figure}[tp]
  \centering
\includegraphics[height = 7cm]{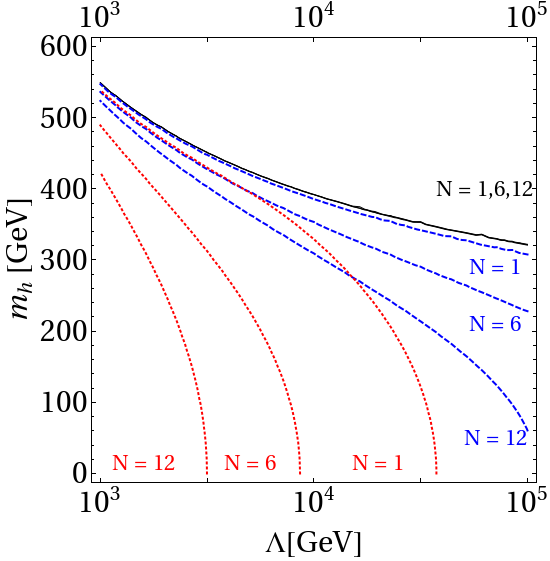}
  \caption{The triviality upper bound on the Higgs mass
as a function of the cut-off $\Lambda$ for $\lambda_{\varphi} (m_{W}) = 0.1$,
for $\lambda_{x \, 0} = 0.1, 1, 2$ 
(black, blue and red curves, respectively) and $N=1,6,12$ (starting with the 
uppermost curve); the region above each 
  curve is excluded the  by the triviality constraint for the 
corresponding set of parameters. 
For $\lambda_{x \, 0} = 0.1$ non-standard effects 
  are so small that
  upper boundaries for $N=1,6$ and $12$ are indistinguishable from the SM, 
accordingly the  black curves  are nearly identical to the bound derived
within the SM.}
  \label{triv_N}
\end{figure}

The triviality bound for $\mh$ as a function of $\Lambda$
is illustrated in Fig.~\ref{triv_N}. Note that the allowed region shrinks as $\lambda_{x \, 0}$ grows
for fixed $N$, and as $N$ increases for fixed $\lambda_{x \, 0}$,
lowering the upper bound on $\mh$ in either case. This behavior is a direct consequence of  
the term $2 N\lx^2$ in (\ref{diff_set1}): increasing $N$ and/or $\lx$ amplifies the evolution of $\lh$. For a given  $(\Lambda,m_{h})$ and fixed $\lambda _{\varphi \, 0}$ there is a range of $\lambda_{x \, 0}$ for which the Landau pole occurs above $\Lambda$, as shown in Fig.~\ref{lambdaMAX}. Note the asymmetry of the allowed (inner) region, which is a consequence of the $8\lx^2$ term in (\ref{diff_set2}). 

\begin{figure}[tp]
  \centering
\includegraphics[height=7 cm]{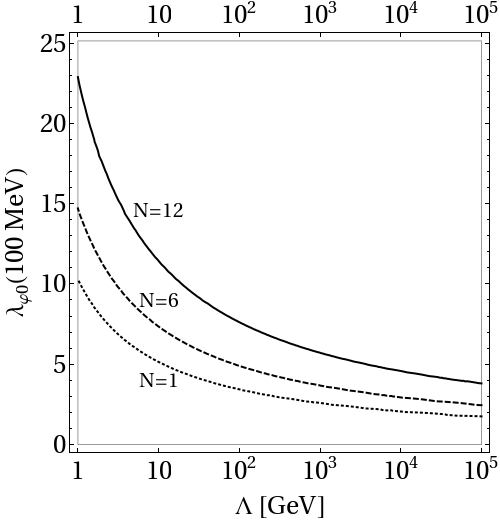}
\includegraphics[height=7 cm]{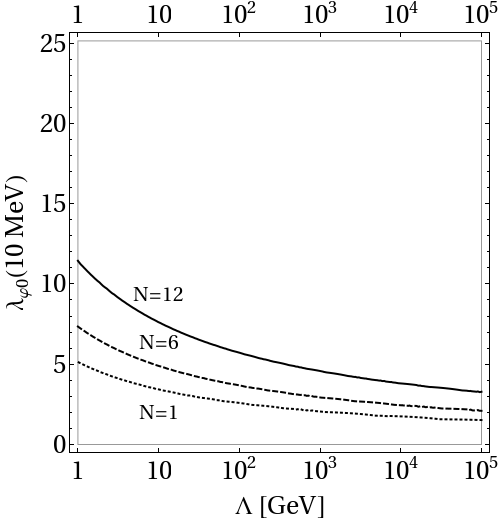}\\
  \caption{Maximum $\lambda_{\varphi \, 0}$ allowed by the triviality condition as a function of  
  $\Lambda$ for $N = 1, 6, 12$ and $\lambda_{x \, 0}= 0$. The right (left)
panel is for initial conditions provided at a $ 10 $ MeV  
($100$ MeV)}
\label{triv_phi}
\end{figure}

\begin{figure}[tp]
  \centering
\includegraphics[height=7 cm]{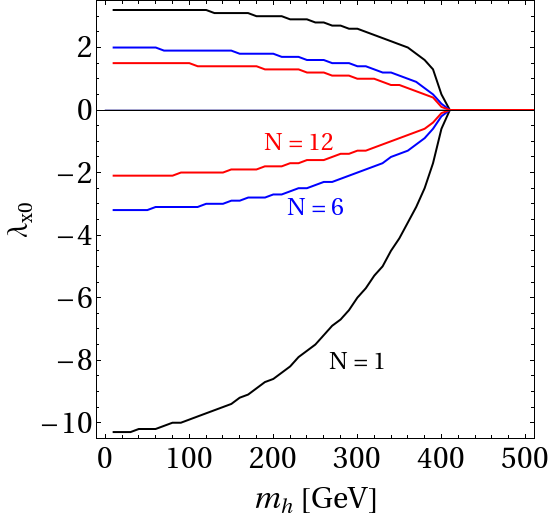}
  \caption{Maximum and minimum $\lambda_{x \, 0}$ allowed by the triviality bound for 
  $\Lambda = 10$ TeV, $ \lambda_{\varphi \, 0} = 0.1 $ and $N = 1, 6, 12$.}
  \label{lambdaMAX}
\end{figure}

In Fig.\ref{triv_phi} we show triviality limits for $\lambda_\varphi$ as a function of $\Lambda$ for the case $\lambda_{x\, 0}=0$ when the $\vp$ evolution decouples from the SM. 

%
\section{Fine-tuning in Higgs mass corrections}
\label{Sec:fin-tun}

If we assume that the SM is an effective theory valid 
at scales below the UV cutoff  $ \Lambda $,
radiative corrections ~\cite{Veltman:1980mj} 
shift the Higgs mass to a scale of order
$ \Lambda/\sqrt{4\pi} $. The 
existing bounds on the scale of new physics imply $\Lambda > O(\tev) $,
which is in conflict with a Higgs mass below $1$ 
TeV range as required by the unitarity constraint \cite{Lee:1977eg}.
A  way of alleviating this conflict is to arrange cancellations to occur 
within the $O(\Lambda^2) $ contributions to $ \mh^2 $. These cancellations
can be obtained through a fine tuning of the parameters or through a
an extension of the SM together with an appropriate new symmetry
(as is the case in supersymmetric models).

The first solution to this problem was proposed by Veltman
~\cite{Veltman:1980mj} who observed that the  1-loop 
quadratic corrections to $ \mh$ would vanish if $m_{h}^2 +m_{Z}^2 + 2 m_{W}^2 - 4m_{t}^2 = 0 $, but this possibility is now experimentally disallowed. Here we shall follow a similar approach 
within the multi-singlet extension of the SM.

In pursuing this approach it is important to determine the possible 
effects generated by higher order radiative corrections.
In a theory  with many couplings $\lambda_{i}$
the general form of leading higher-order contributions (those
containing the highest power of $\log(\Lambda)$) to the
quadratically divergent contributions to $ \mh^2 $ 
at $n+1$ loops take the form~\cite{Einhorn:1992um}
\begin{eqnarray}
\delta \mh^2 = \Lambda^2 \sum_{n=0}^{\infty} f_{n} (\lambda_{i}) \left[
\log \left( \frac{\Lambda}{\mu} \right) \right]^n
\label{eq:ein}
\end{eqnarray}
where $\mu$ is the renormalization scale and the 
coefficients $f_{n}$ satisfy
\begin{eqnarray}
(n+1)f_{n+1} = \mu \frac{\partial}{\partial \mu} f_{n} = \beta_{i} \frac{\partial}{\partial \lambda_{i}} f_{n}\,.
\label{recursion}
\end{eqnarray}

Thus in order to determine the two-loop leading logarithmic 
 corrections to the Higgs mass one needs the one-loop 
corrections and the first order beta functions 
(for generic results and applications to the single scalar case see
\cite{OD-MSc-thesis}). The results for $N$-singlet model read: 
 \begin{eqnarray} 
\delta m^2_{h \,1-loop} &=& \frac{\Lambda ^2}{16 \pi^2} \left(
12 \lh + 2 N \lambda_x - 12 y_t^2 + \frac{3}{2} g_1^2 + \frac{9}{2} g_2^2 \right) \nonumber\\
& & - \frac{1}{16 \pi^2} \left[
6 \lh m_{h}^2 \, \log   \left( \frac{m_{h}^2+\Lambda^2}{m_{h}^2}  \right) +
2 \lx m_{\varphi}^2 \, \log   \left( \frac{m_{\varphi}^2+\Lambda^2}{m_{\varphi}^2}  \right)
\right]
\label{1-loop}\\
\delta m^2_{h \,2-loops} &=& \frac{\Lambda ^2}{(16 \pi ^2)^2} \, \log   \left( \frac{\Lambda}{\mu} \right)
\left[ 25 g_1^4 + 9 g_1^2 g_2^2 - 15 g_2^4 + 34 g_1^2 y_t^2 + 54 g_2^2 y_t^2 \right. \nonumber\\
& & + 
 192 g_3^2 y_t^2 - 180 y_t^4  \lh - 36 g_1^2 - 108 g_2^2 \lh + 
 144 y_t^2 \lh + 288 \lh^2 \nonumber\\
& & \left. - 3 N g_1^2 \lx - 
 9 N g_2^2 \lx + 12 N y_t^2 \lx + 24 N \lh \lx + 
 40 N \lx^2 + 2 N \lx \lvp \right]
\label{2-loop}
\end{eqnarray} 
the logarithmic terms ($\propto \mvp^2 \log \Lambda$) in the one-loop correction were kept since they are relevant 
in the range $\mh \ll \mvp \lsim \Lambda$); terms $\propto \mh^2 \log \Lambda$ are always numerically negligible and are included for completeness. 
For the numerical results the renormalization scale was chosen to be the vacuum expectation value of the Higgs field, $\mu = 246\gev$. The SM result can be recovered in the limit $\lambda_x = \lambda_{\varphi} = 0$. 

\begin{figure}[h!]
\centering
\includegraphics[height = 6.5 cm]{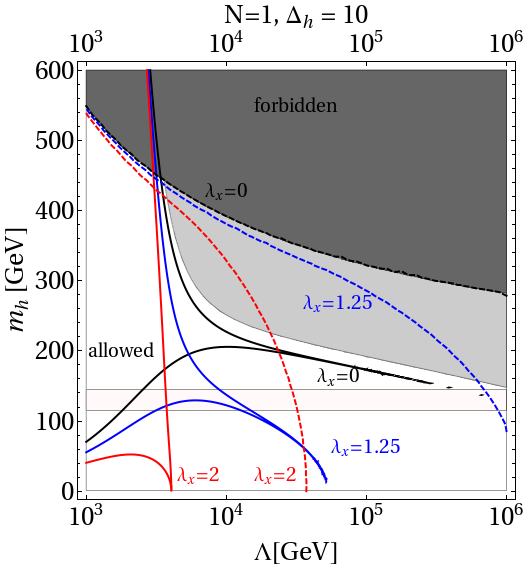}
\includegraphics[height = 6.5 cm]{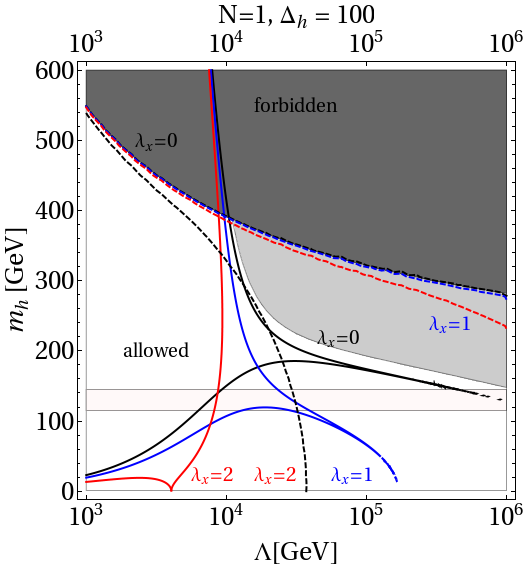}\\
\includegraphics[height = 6.5 cm]{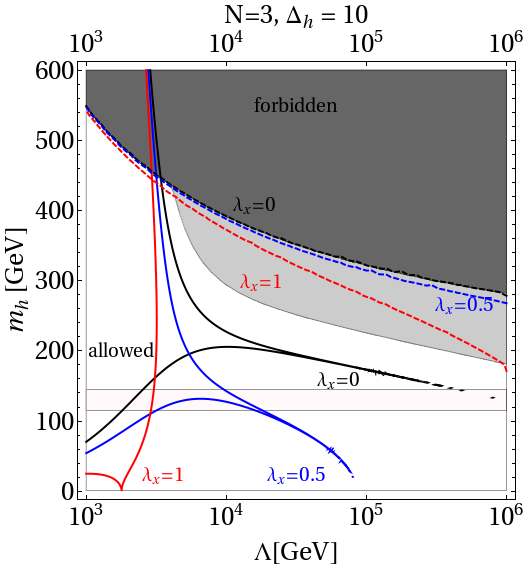}
\includegraphics[height = 6.5 cm]{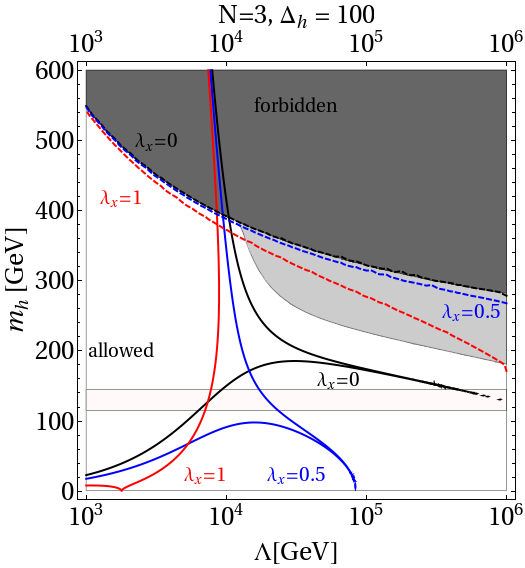}
\caption{Constraints on $\mh$ and $\Lambda$
for $ \Delta_h=1,10$ and $ N=1,3$. The
horizontal band shows the presently allowed range for the
Higgs-boson mass~\cite{H-mass-limits}: $115 \gev < \mh < 141\gev$.
Dashed curves correspond to the triviality limits
for the SM ($\lx=0$, dashed black curve) 
  and singlet extensions for several values
of $ \lx $ (dashed blue and red curves); regions below 
the curves are allowed by this constraint. 
  The  solid lines correspond to the two-loop fine-tuning  constraint
for the SM  ($\lx=0$, solid black) and singlet extensions for 
several values of $ \lx $ (solid red and blue curves), 
the regions to the left of the corresponding funnels are
allowed by this constraint. 
Every point inside the dark (light) gray region is forbidden by the triviality
(find tuning) condition for any value of $\lx$. 
For every allowed point outside the dark gray and gray regions there exist $|\lx| < 8\pi$, $10\gev < \mvp < 10^4 \gev$ and $0 < \lvp< 2$ such that (\ref{delta}) is fulfilled.
}
\label{FT}
\end{figure}

\begin{figure}[h!]
\centering
\includegraphics[height = 6.5 cm]{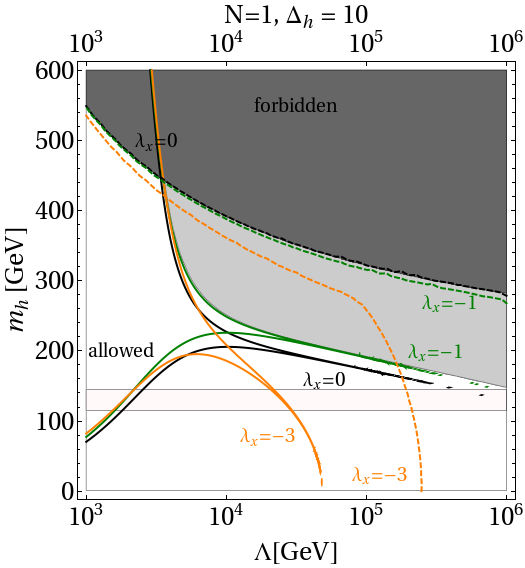}
\includegraphics[height = 6.5 cm]{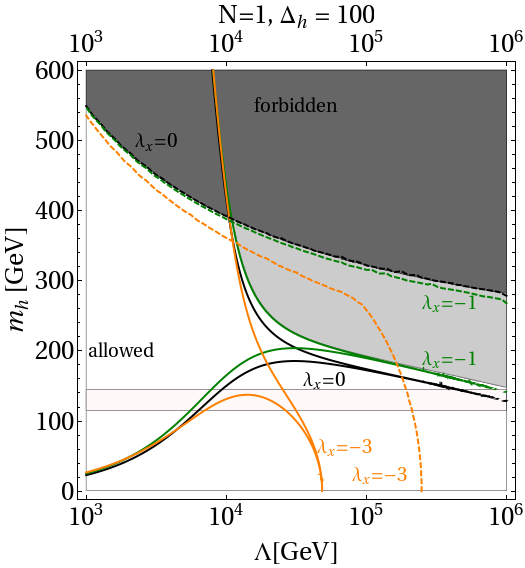}\\
\includegraphics[height = 6.5 cm]{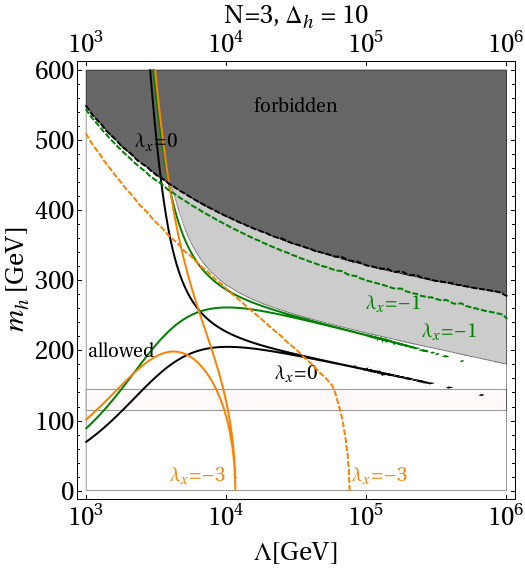}
\includegraphics[height = 6.5 cm]{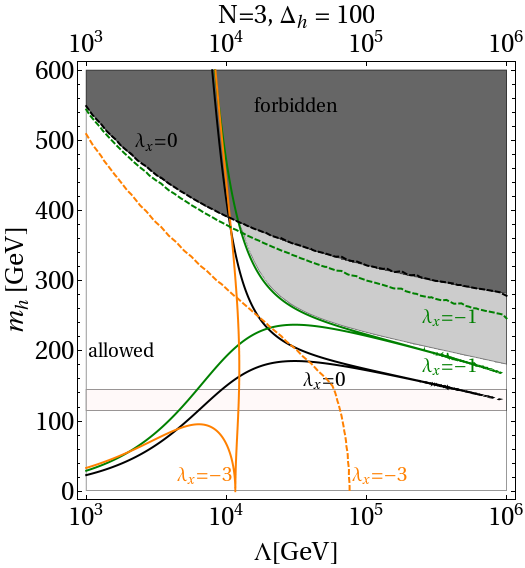}
\caption{Similar as Fig.~\ref{FT} for negative $\lx$.
}
 \label{FT_neglx}
\end{figure}

We implement the fine tuning condition in a standard manner~\cite{Kolda:2000wi}   by requiring
\begin{equation}
\left|\frac{ \delta \mh^2 }{ m_{h}^2}\right|  \equiv  \left|\frac{\delta m^2_{h \,1-loop} + \delta m^2_{h \,2-loops} }{ m_{h}^2}\right| \leq \Delta_{h} 
\label{delta}
\end{equation}
and determine the region in the $(\Lambda , \mh )$ plane where this is
obeyed for a given choice of the fine tuning parameter $\Delta_h$. 
Specifically, for each choice of $(\Lambda , \mh )$ 
with fixed $N$ and $\lx$ we check if there exist $m_{\varphi}$  and $\lvp$  
such that condition (\ref{delta}) is satisfied; the results are presented
in Figs.~\ref{FT} and \ref{FT_neglx}.

It is worth noting that the two-loop  $\vp$ corrections to $ \mh$
(\ref{2-loop}) grow as $\lx/(16 \pi^2) \ln(\Lambda/\mu)$ relative to the one-loop 
contributions(\ref{1-loop}).
Perturbative consistency then requires this factor to be $ \le 1 $, which
translates into a maximal cutoff $\Lambda_{\rm max}$ beyond which
perturbation theory is no valid:
\beq
\Lambda \lsim \Lambda_{\rm max} = \mu e^{4 \pi^2/(5\lx)}
\label{Lam_max}
\eeq
Therefore the regions of large $\Lambda$ in Figs.~\ref{FT} and \ref{FT_neglx} should be considered
with a certain caution (that should increase with $\lx$). The range of $\Lambda$ is extended 
up to $10^6\gev$ only for the purpose of the SM ($\lx=0$) and small $\lx$ cases 
($ \lx \lsim 0.95$ for $ \mu = 246 \gev $).

We now consider the regions in the $(\Lambda , \mh )$ excluded by the triviality
and fine tuning constraints, first for $ \lx \ge 0 $ and then for $ \lx < 0 $.

\paragraph{$\lx \ge 0 $.}
 As one can see in Fig.~\ref{FT}, the fine-tuning condition
(\ref{delta}) defines a 
region in the $(\Lambda , \mh )$ plane with a 
characteristic funnel-shaped boundary (bounded, for various values of $\lx$, 
by the solid lines shown in the figure). Given the experimental bounds~\cite{H-mass-limits} on $\mh $, 
$115\gev < \mh < 141\gev$, we see that 
as $ \lx$ or $N$ grows, the  lower branch of the funnel 
allows larger regions of $\Lambda$ for fixed $\Delta$,
or smaller values of $\Delta_h$ for fixed $\Lambda $; in contrast
the upper branch of the funnel allows smaller regions of $\Lambda $
 for fixed $\Delta_h$, or larger values of $\Delta_h$ for fixed $\Lambda $.
As a result the combined effect is complicated, for example the 
region allowed by (\ref{delta}) and the experimental
limits on $ \mh $ shrinks as $\lx$  goes from $1.25$ to $2$ when 
$N=1$, and as $\lx$ goes from $0.5$ to $1$ for $N=3$.
For fixed $\lx$ and $N$ the allowed range of $\Lambda$ 
shrinks as $\Delta_h$ becomes smaller.

Several comments are in order here. 
For the SM ($\lx=0$) there is a disallowed region for 
$\Lambda$ between $1-10$ TeV and $\sim 10^3$ TeV;
this region shrinks as $\Delta$ grows.
For instance, if $\Lambda \sim 10^4 \gev$ 
then the present limits on the Higgs boson mass,  
can be accommodated only for large fine tuning: $\Delta_h > 100$.  
In order to have $ \Delta_h < 100  $ a modification of the SM
must be introduced, or low ($\sim1 \tev $) or very high $ \gsim 10^3 \tev$
values of $ \Lambda $ must be assumed.

Note that for each point $(\Lambda,\mh)$ below the SM ($\lx=0$) upper branch there exist $\lx$ such, 
that the point  is allowed. Therefore in the model we discuss here, 
for a fixed cutoff $\Lambda$, much smaller Higgs-boson masses are allowed. In particular 
the range of $115\gev < \mh < 141\gev$ could be easily accommodated with a cutoff much larger than
the SM one.


Let's concentrate on the cutoff $\Lambda = 10^4\gev$. As it is seen from the left panels is Fig.~\ref{FT} for such $\Lambda$, 
in order to have the range of presently allowed Higgs masses $115 \gev < \mh < 141\gev$ within the region of fine tuning $\delta\mh^2/\mh^2$ below 10, 
one needs $\lx$ of the order of $1.2-1.3$ and $0.5-0.6$ for
$N=1$ and $N=3$, respectively. However, if we relax the fine tuning condition so that $\delta\mh^2/\mh^2 < 100$ 
(see the right panels of Fig.~\ref{FT}) then, as expected, lower $\lx$ is sufficient; $\lx\simeq 1$ for $115\gev < \mh < 130\gev$ and $\lx \simeq 0.5$ for
$130 \gev < \mh < 141 \gev$, for $N=1$. If $N=3$, then $\lx\simeq 0.5$ is large enough to bring the whole $\mh$ range into
the region of the  fine tuning $\delta\mh^2/\mh^2 < 100$. 
In general,  $\lx = O(1) $ is needed to ameliorate the SM fine tuning.

On the other hand, one should remember that for each choice of $ \mvp $,
$\lx $ is restricted  from above by the condition (\ref{mphi_bound}), therefore if $0.5 < \lx < 1.5$, then the corresponding $\vp$ mass must be in the range $174\gev < \mvp < 301\gev$. 
Alternatively, if we require  $\mvp = 15 - 60 \gev$ (the range relevant 
for direct DM detection, see Sec.~\ref{Sec:dir-det}) then 
 $0.06 < \lx < 0.24$; as seen from Fig.~\ref{FT} such low $\vp$ masses 
are inconsistent with $ \Delta_h =100,~ \Lambda=10^4\gev$ and $N=1$ or $3$. 
In order to allow for that low mass, either larger $N$, $ \Delta_h > 100$ 
or lower cutoff $\Lambda$  would be necessary.

It is worth noting that for large $\mh$ ($\mh \gsim 310\gev$) the SM contribution to the quadratic divergence is dominated by the Higgs boson, therefore in order to enlarge the allowed region the extra contribution from $\vp$ in the loop should be of the opposite sign
(corresponding to $\lambda_x <0$).

\paragraph{$\lx < 0 $.}
In this case the limit (\ref{mphi_bound}) provides no restriction, 
on the other hand the vacuum stability condition (\ref{stab_con}) becomes important. 
This constraint allows  $|\lx| >1 $ but only for $ \mh > 170 \gev $ due to
the unitarity limit (\ref{unit_con});
saturating this limit by taking $\lvp=8 \pi$ requires
 $|\lx| < 0.65 - 0.85$ for $110 \gev < \mh < 141\gev$.
In this range the allowed values of $ \Lambda $ are more restricted
when the singlets are included than for the SM, this is because for
$\lx<0$ the 1-loop $\vp$-contribution to
$\delta \mh^2$ is of the same sign as the top-quark one, therefore for small $\lx$ funnels are shifted towards larger values of $ \mh $, as needed
by (\ref{delta}) in order to compensate the top and $\vp$ effects;
as $\lx$ increases the
funnels move downwards as a consequence of the growing 
2-loop contributions to $\delta m^2_{h \,2-loops}$.

\medskip 

The dashed curves in 
Figs.~\ref{FT} and \ref{FT_neglx} give the triviality limits 
on $ \mh $ for various values of $ \lx $
(see also Fig.~\ref{triv_N}). As seen 
from the figures, this constraint is not important for $115 \gev < \mh < 141\gev$.

Summarizing this section, one can say, that for $0 < \lx \lsim 1$ 
the model allows to shift the UV cutoff up to $\Lambda \lsim 10^4\gev$
keeping the Higgs-boson mass within the experimentally allowed range $115 \gev < \mh < 141\gev$.
The case $\lx<0$ is disfavored, as the cutoff in this case 
appears to be lower than in the SM as long as we stay
within the range of parameters allowed by perturbative expansion. 

\section{DM relic abundance}
\label{Sec:DM}

The singlet  $\vp$, being odd under the $Z_2$ symmetry, is stable, and therefore a dark matter candidate,
in which case the parameters of the theory should also lead to the DM abundance 
consistent with the WMAP observations, $\dm = 0.110 \pm 0.018$ ($3 \sigma$) \cite{Nakamura:2010zzi}. In order to gauge the consequences of this constraint on our model we first
calculate the relic DM abundance by solving the appropriate Boltzmann equation.
Considering first the single component ($N=1$) case, we have \cite{kolb}:
\begin{eqnarray}
 \frac{d f}{d T}  = \frac{ \sv}{K} (f^2 - f_{EQ}^2) , \lsp  K(T) =\sqrt{ \frac{4 \pi^3 g_\star(T)}{45 m_{Pl}^2} }, 
 \lsp f_{EQ}(T) = \int \frac{d^3p}{(2\pi)^3} \frac{1}{\exp(E/T) \pm 1)} 
 \label{b_eq}
\end{eqnarray}
where $f\equiv n/T^3$,  $n$ is the particle number density of the DM candidate (in our case the singlet), $f_{EQ}(T)$ is the equilibrium distribution, $g_\star(T)$ is the number of relativistic degrees of freedom at temperature $T$, $m_{Pl}$ is the Planck mass and $\sv$ is the thermally-averaged cross section for $DM + DM \rightarrow SM + SM$ annihilation processes (\cite{Gondolo:1990dk}, \cite{Guo:2010hq}):
\begin{eqnarray}
& &\sv = \frac{x}{16 m_{\varphi}^5 K_{2}^2(x)}
\int_{4m_{\varphi}^2}^{\infty} ds K_{1}\left(\frac{\sqrt{s}}{T}\right) \sqrt{s-4m_{\varphi}^2} \, \, \hat{\sigma}(s) , \\
& & \hat{\sigma}(s) = 2 \sqrt{s(s-4m_{\varphi}^2)} \,  \sigma (s)
\end{eqnarray}
where $x=m_\varphi/T$, $K_{1,2}(x)$ are the modified Bessel functions, and $\sigma(s)$ is the $DM + DM \rightarrow SM + SM$ annihilation cross section normalized in the standard manner. In analytical
estimates is useful to note that $\sv \sim  \hat{\sigma} (4 m_{\varphi}^2)/ 4 m_{\varphi}^2 $
for non-relativistic DM; however we do not use this approximation in the numerical results shown below.

In the case of $N$-singlet scalar DM candidates with $O(N)$ symmetry, all components  contribute equally to the relic abundance, just like a single field with $N$ degrees of freedom. Therefore the total abundance equals
\begin{eqnarray}
\Omega_{\mathrm{DM}}^N = \sum_{i} \Omega_{\mathrm{DM}}^{i} = N  \Omega_{\mathrm{DM}}^{1}
\label{omega_N}
\end{eqnarray}
where $\Omega_{\mathrm{DM}}^{i}$ is the dark matter relic density from the $i$-th scalar field. By virtue of
the $O(N)$ symmetry it is sufficient to consider the Boltzmann equation for one of the components 
of $\vp$. 

In the following we will discuss two different limiting cases for the Boltzmann equation: Cold Dark Matter (CDM) and Feebly Interacting Dark Matter (FIDM). We do not consider the hot dark matter solution, since it
is inconsistent with structure formation at the galaxy scale, see e.g. \cite{896695}.

\begin{figure}[h]
\includegraphics[height = 2.6 cm, width = 2.6 cm]{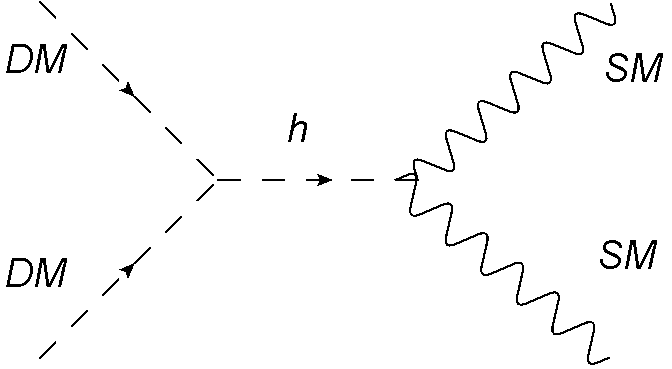}\hspace*{.4cm}
\includegraphics[height = 2.6 cm, width = 2.6 cm]{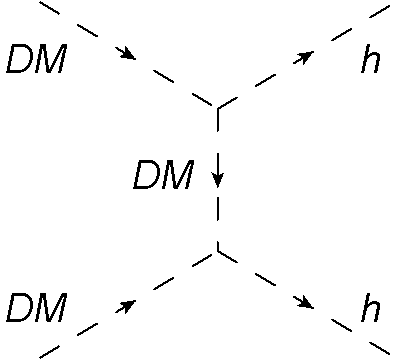}\hspace*{.4cm}
\includegraphics[height = 2.6 cm, width = 2.6 cm]{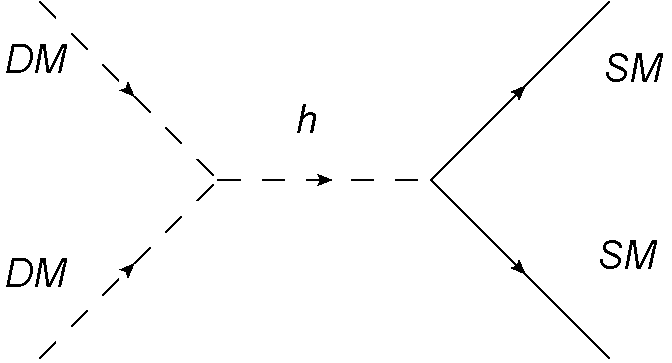}\hspace*{.4cm}
\includegraphics[height = 2.6 cm, width = 2.6 cm]{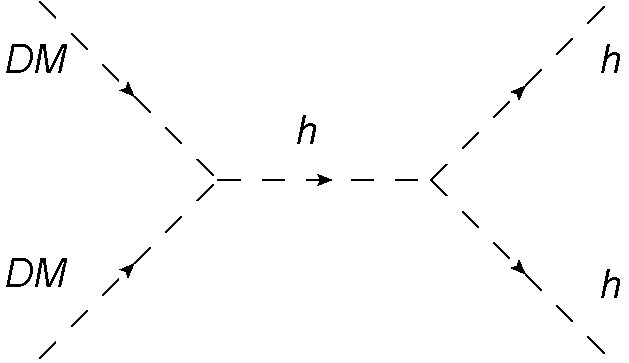}\hspace*{.4cm}
\includegraphics[height = 2.6 cm, width = 2.6 cm]{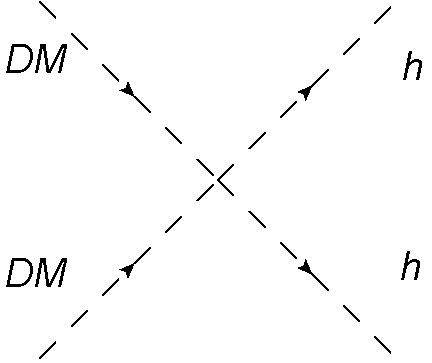}
  \caption{Feynman diagrams illustrating $\vp \vp$ annihilation into SM particles.}
\label{ann_diag}
\end{figure}

The diagrams contributing to $\vp \vp$ annihilation into SM particles are shown in Fig.~\ref{ann_diag}.
The corresponding cross sections (for $N=1$) are available in the literature
(e.g. \cite{Burgess:2000yq} and \cite{Guo:2010hq}); we have verified the results of \cite{Guo:2010hq}:
\begin{eqnarray}
\hat{\sigma}_{WW}(s) &=& \frac{\lambda_{x}^2 }{2\pi} \sqrt{1 - \frac{4 M_{W}^2}{s}}   \frac{ s^2 }{(s-m_{h}^2)^2+\mh^2 \Gamma_{h}^2} 
\left( \frac{12 M_W^4}{s^2} - \frac{4 M_W^2}{s} + 1 \right) \cr
\hat{\sigma}_{ZZ}(s) &=& \frac{\lambda_{x}^2 }{4\pi} \sqrt{1 - \frac{4 M_{Z}^2}{s}}   \frac{ s^2 }{(s-m_{h}^2)^2+\mh^2 \Gamma_{h}^2} 
\left( \frac{12 M_Z^4}{s^2} - \frac{4 M_Z^2}{s} + 1 \right) \cr
\hat{\sigma}_{\overline{f}f}(s) &=& \frac{\lambda_{x}^2}{\pi} \left(\sqrt{1 - \frac{4  m_{f}^2}{s}} \, \right)^3
 \frac{ m_{f}^2 \, s }{(s-m_{h}^2)^2+\mh^2 \Gamma_{h}^2} \cr
\hat{\sigma}_{hh}(s) &=&
\frac{\lambda_{x}^2 }{4 \pi}\sqrt{1 - \frac{4 m_{h}^2}{s}} \left(
 \frac{(s + 2m_h^2)^2}{(s-m_h^2)^2}
+ \frac{32 v^4 \lambda_{x}^2}{(s-2m_h^2)^2} \left(\frac{1}{1-\xi^2} + F(\xi)
\right)
-\frac{16 v^2 \lambda_{x}(s+2m_h^2)}{(s-2m_h^2)(s-m_h^2)} F(\xi)
\right)
\label{xsec1}
\end{eqnarray}
where $F(\xi) = \mathrm{ArcTanh}(\xi)/\xi$, $\xi = \sqrt{(s-4\mh^2)(s-4\mvp^2)}/(s-2\mh^2)$. 
The total cross section is then
\beq
\hat{\sigma} (s) = \hat{\sigma}_{WW}(s) + \hat{\sigma}_{ZZ}(s)  + \sum_{f} \hat{\sigma}_{\overline{f}f}(s) +\hat{\sigma}_{hh}(s) \label{xsec}\\
\eeq
where the sum runs over all fermions $f$. 

When $ N>1$ one has to modify the Higgs width as it can decay into $N$ possible $\varphi_i \varphi_i$ final states:
\begin{eqnarray}
\Gamma_{h} = \Gamma_{h \rightarrow SM} +  \Gamma_{h \rightarrow \varphi \varphi} 
\label{h_width}
\end{eqnarray}
\begin{eqnarray}
 \Gamma_{h \rightarrow \varphi \varphi} = \frac{N v^2 \lambda_{x}^2 }{8 \pi \mh^2} \sqrt{\mh^2-4m_{\varphi}^2} 
\, \theta_{H}(\mh - 2 m_{\varphi})
\end{eqnarray}
where $\theta_{H}$ is the Heaviside step function.

\subsection{Cold Dark Matter }
\label{Subsec:CDM}

In the CDM approximation the solution to the Boltzmann equation for the $O(N)$ model yields
the relic density~\cite{kolb}
\begin{eqnarray}
\Omega_{\mathrm{DM}}^N h^2 = N \frac{\rho_{\mathrm{DM}}^1}{\rho_{\mathrm{crit}}}=1.06 \times 10^9 
\frac{N x_f}{\sqrt{g_{*}} \, m_{Pl} \sv } \frac{1}{\mathrm{GeV}}
\label{approx_cdm}
\end{eqnarray}
where $x_f  \equiv m_{\varphi}/T_f$, and $T_f$ is the freeze-out temperature given in the first approximation by
\begin{eqnarray}
x_f  = \log  \left( 0.038 \frac{\sv  m_{Pl} \, m_{\varphi} }{\sqrt{g_{*}(T_f) x_f }} \right)
\label{x_f}
\end{eqnarray}
where $\sv$ is the thermally averaged cross section for $\varphi_{1} + \varphi_{1} \rightarrow SM + SM$. 
Note that the $N$-dependence of $ x_f $ enters through the Higgs width (\ref{h_width}) 
contained in $ \sv $; this weak dependence is furthered softened by the $\log$ function 
in (\ref{x_f}). Therefore it is reasonable to neglect the $N$ dependence in $ x_f$ 
(at least in the parameter range where the perturbative expansion is valid).
Note also that in the case where the $t$- and $u$-channel annihilation diagrams with $\vp$ exchange can
be neglected (e.g. when $\mvp \gg \mh$ and/or if $\lx \ll 1$), the abundance depends only
on the combination $ N \lx^2$.

\begin{figure}
  \centering
  \includegraphics[height = 5.1 cm]{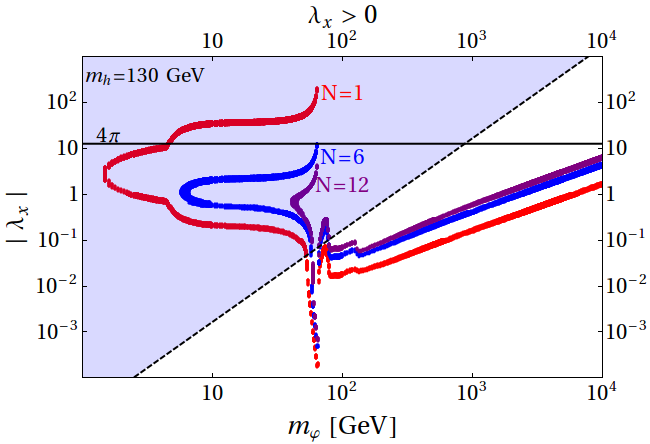}
  \includegraphics[height = 5.1 cm]{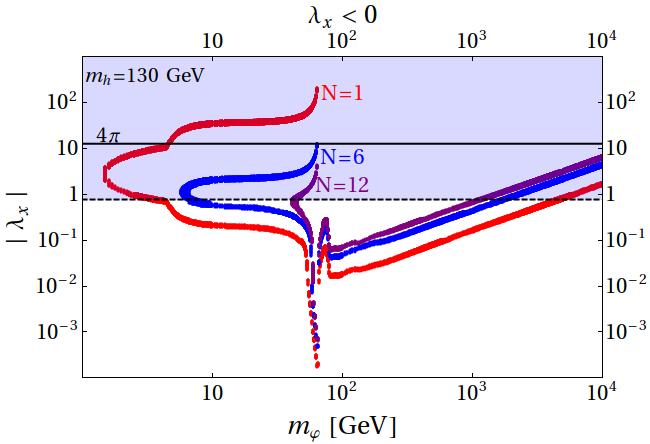}
  \caption{
  The coupling $\lx$ as a function of $\mvp$ obtained from the requirement (\ref{dm}); 
for a the Higgs $\mh=130\gev$, $N=1,6$ and $12$ (red, blue and purple bands respectively),
and  $\lx>0$ (left panel) or $\lx<0$ (right panel). The blue areas in the left and right panels 
correspond to the regions disallowed by the consistency condition (\ref{mphi_bound}) 
and  stability constraint (\ref{stab_con}) for  $\lvp=8\pi$, respectively.
 The thick black lines show the unitarity limit (\ref{unit_con}) saturated by $|\lx|=4\pi$.
 }
\label{cdm_sol_130}
\end{figure}

For a given choice of $N$ and $(\mh,m_{\varphi})$ we will look for $\lambda_x$ 
such that the constraint $\dm = 0.110 \pm 0.018$ is satisfied:
\beq
0.092 < \Omega_{\mathrm{DM}}^N < 0.128 \,.
\label{dm}
\eeq
Even though (\ref{approx_cdm}) often leads to accurate results for the abundance, 
in deriving our numerical results we used 
micrOMEGAs~\cite{Belanger:2008sj} -- a code dedicated for calculation of DM properties. 
Our results are presented in Fig.~\ref{cdm_sol_130}, where we have also included
the restrictions derived form the consistency condition (\ref{mphi_bound}) and
the vacuum stability  constraint (\ref{stab_con}), which are
relevant for $ \lx > 0 $ and $\lx<0$, respectively.

To discuss solutions for $\lambda_x$ shown in Fig.~\ref{cdm_sol_130} it is useful to consider the following regions for $(\mh,m_{\varphi})$:
\bit
\item $m_{\varphi} \gsim \mh$ \\
In this region $\mvp$ is large, so the $t$- and $u$-channel annihilation diagrams with $\vp$ exchange 
can be neglected. In this approximation $\sv $ depends on $\lx^2$, so there is no difference between
left and right panels in Fig.~\ref{cdm_sol_130}. For $\lx>0$, the constraint  (\ref{dm}) is consistent with
(\ref{mphi_bound}) in this mass range, however for negative $\lx$ (\ref{stab_con}) 
requires $  m_\varphi \lsim 1 \tev $. 

For $\lx > 0$  the range of very heavy scalar masses
$\mvp \sim 1-8\tev$, corresponds to $\lx\sim 1-10$, which though large, remains below
the unitarity and consistency limits. In this region of parameter space the 
singlets provide a substantial contribution to $\delta \mh^2$ that can ameliorate the hierarchy problem (see  Fig.~\ref{FT}), and was utilized for this purpose in \cite{Grzadkowski:2009mj}.
\item $ \mh/2 \lsim m_{\varphi} \lsim  \mh$ \\
In this mass range the interference between $s$- and $t$- and $u$-channel annihilation diagrams is relevant, so some differences between negative and positive $\lx$ solutions are visible.  For large $N$ and $ \lx > 0 $  
a small region is excluded  by  (\ref{mphi_bound}).
\item $\mvp \sim \mh / 2$ \\
In the vicinity of the resonance, the annihilation cross section is strongly enhanced,
so that (\ref{dm}) can be satisfied only for small $ |\lx| $. 
\item $m_{\varphi} \lsim  \mh / 2$ \\
In this case the process $\varphi\varphi\to h h$ is kinematically forbidden,
so only the  $s$-channel Higgs exchange graphs with no Higgs-bosons
in the final state contribute; as a consequence $ \sv \propto \lx^2$
and there is no difference between left and right panels Fig.~\ref{cdm_sol_130} for this mass range. 
For $\lx >0$ this mass region is  excluded by (\ref{mphi_bound}), 
while for $\lx < 0$ values of $\mvp$ below $\mh/2$ are allowed depending on $N$.
\eit

For scalar masses below $\sim 1\gev$ there are no CDM solutions.

\subsection{Feebly Interacting Dark Matter (FIDM)}
\label{Sec:FIDM}

In order to thermalize with the SM particles, any other species 
of density $n$ must 
have a thermalization rate $\Gamma = n \sv $ larger than the expansion rate
$H$. For relativistic singlets one can estimate $ \Gamma
\sim \lx^2 T /(8 \pi) $ while $H \sim T^2 / m_{Pl}$, therefore equilibrium 
prevails for $ T < T_{EQ} $ where
\beq
T_{EQ}  \sim \frac{m_{Pl} \lambda_{x}^2}{8 \pi}\,.
\label{teq}
\eeq 
If $\lx$ is very small ($\sim 10^{-9}$) the species of DM 
being considered here does not equilibrium until very late times 
$T_{EQ} \sim 1$ GeV;
for $ \lx \lsim 7 \times 10^{-16} $, $T_{EQ} < 2.7\,^oK $ and
the singlets would not have equilibrated before the present epoch. 

In the following we will consider the case 
of feebly interacting DM (FIDM).
We will assume that the DM number density $f$ was negligible at 
the Big Bang: $\lim_{T \rightarrow \infty}{f(T)} = 0$ and
solve the Boltzmann equation for $N=1$
(\ref{b_eq}) with this initial condition~\footnote{In practice,
the Boltzmann equation (\ref{b_eq}) simplifies for this initial condition since
the $f^2$ term on the right-hand side can be dropped: the DM particles do 
not annihilate when $\lim_{T \rightarrow \infty}{f(T)} = 0$.}; this then 
determines the relic abundance:
\begin{eqnarray}
\Omega_{\mathrm{DM}}^N h^2 = N \frac{\mvp n }{\rho_{crit}} = 
\frac{N\mvp T_{\gamma}^3}{\rho_{crit}} f
\label{om-fidm}
\end{eqnarray}
where $T_{\gamma}$ is the present photon temperature, 
and $\rho_{crit}$ is the critical density.  Following this
procedure it
should be remembered that the electroweak phase transition occurs at $T_{EW} \simeq 300$ GeV. Above this temperature, the only tree level contribution to the annihilation cross section comes from the $\lambda_x H^\dagger H \vp^2 $ coupling, therefore 
\begin{eqnarray}
\hat{\sigma}_{>300} = \frac{\lambda_x^2}{4 \pi}\sqrt{1-\frac{4\mh^2}{s}}
\end{eqnarray}

\begin{figure}
 \includegraphics[height = 5 cm, width = 7.6 cm]{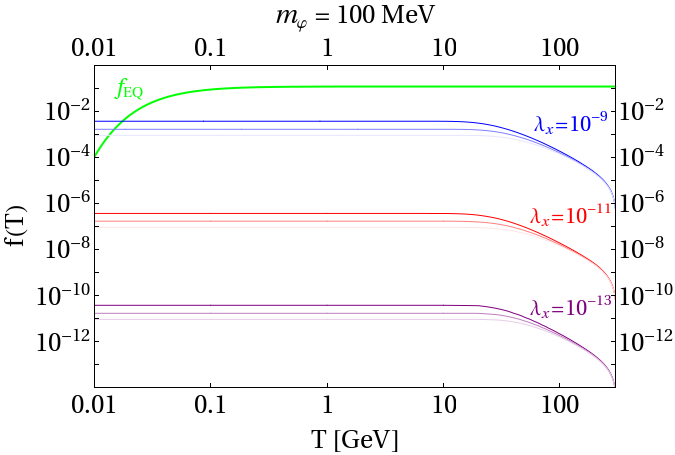}
 \includegraphics[height = 5 cm, width = 7.6 cm]{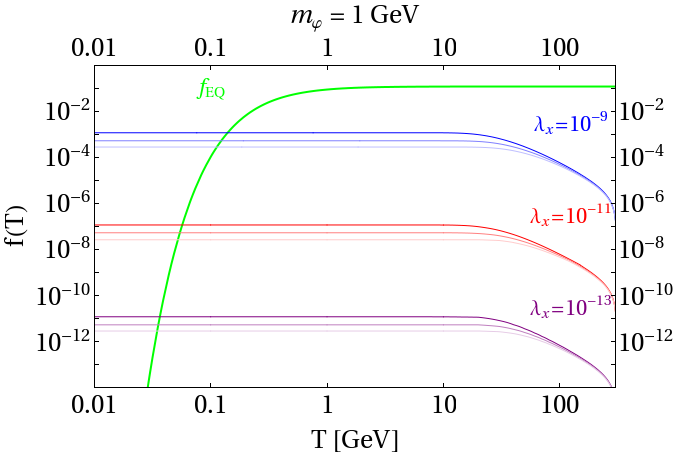} 
 \includegraphics[height = 5 cm, width = 7.6 cm]{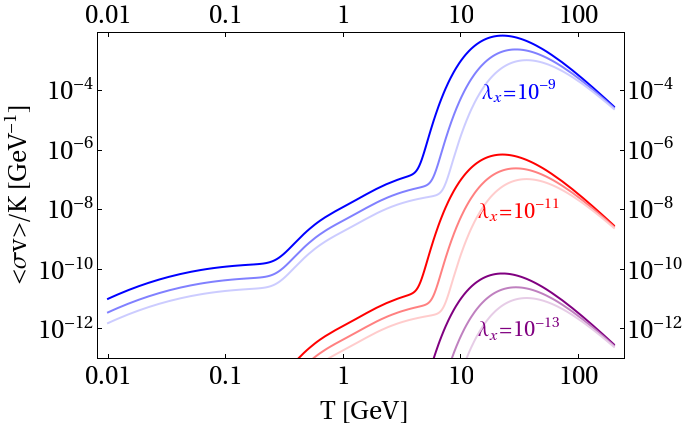} 
 \includegraphics[height = 5 cm, width = 7.6 cm]{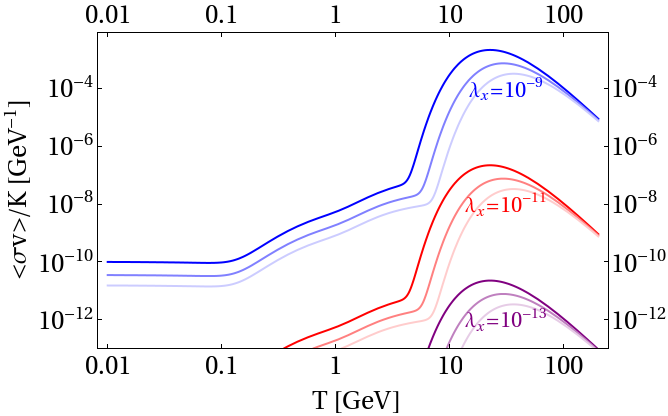} 
 \caption{
Top panels: FIDM solutions to the Boltzmann equation (\ref{b_eq}) with boundary condition $f(T=\infty)=0$, $N=1$ for two values of $\mvp$ and $\mh = 100\gev$ (thin lines), $130\gev$ (medium lines), 
$160\gev$ (thick lines)
 and $|\lambda_x| = 10^{-13}$ (bottom curves), $10^{-11}$ (middle curves), 
 $10^{-9}$ (top curves). The curve labeled $ f_{EQ}$
corresponds to the equilibrium distribution.
Bottom panels: corresponding curves for $ \sv/K $ in (\ref{b_eq}).}
\label{no_equilibrium_f}
\end{figure}

\begin{figure}
\center
 \includegraphics[height = 5 cm]{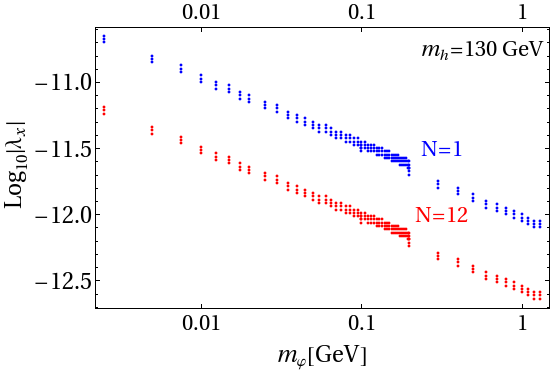}
  \caption{Solutions to the Boltzmann equation for FIDM, for $N=1,12$ scalars and $m_{h} = 130\gev $.}
\label{FIDM_sol}
\end{figure}


The case of FIDM was discussed first
by McDonald in \cite{McDonald:2001vt} and more recently also in \cite{arXiv:0910.2235}
and \cite{Yaguna:2011qn}. In \cite{McDonald:2001vt} the 
contribution from the process $ \vp \vp \leftrightarrow H $ 
was included as a separate, and dominant,
term in the Boltzmann equation. 
Here we obtain similar results following a different approach, 
where the presence of a non-zero width $ \Gamma_H $ 
(see (\ref{xsec1})) accounts for the
creation and decay contributions of the Higgs boson.
In other words, to avoid possibility of double counting we assume 
that the thermal generation of scalars is included
in the processes $DM + DM \leftrightarrow h \leftrightarrow SM+SM$.

The solutions of (\ref{b_eq}) for parameters relevant for insuring
(\ref{dm}) are presented in top panels of Fig.\ref{no_equilibrium_f}. 
As seen from 
this figure, if $\lx \sim 10^{-9 }$ the scalars will be in equilibrium
 with the SM only for  $T\lsim 100\mev$ 
in agreement with the estimate
(\ref{teq}); for smaller $\lx$ 
equilibrium is reached only for even lower temperatures. 
In order to understand the behavior of the
solutions for $f$ we plot in the bottom panels 
of Fig.\ref{no_equilibrium_f} the factor $\sv/K$ 
appearing in (\ref{b_eq})) as a function of $T$. We
see that $\sv/K$ reaches its maximum for $T\sim 20-30\gev$, and is strongly suppressed for $T \lsim 10\gev$. Because of this $f$ becomes $T$-independent
for $T\lsim 20-30\gev$, as observed in the top panels of the figure.

In the case of non-equilibrium solutions of the Boltzmann equation, 
(\ref{dm}) is satisfied only if $\lx \ll 1$. For these small
couplings we have $ \sv \propto \lx^2 $ up to small corrections,
in addition the term $\propto f^2 $ on the right-hand side of (\ref{b_eq})
is subdominant, so that to a good approximation $ f\propto \lx^2 $.
It then follows that for FIDM the relic density
depends on the DM parameters in the combination
$\dm  \sim N \mvp \lx^2 $; this behavior is indeed observed in Fig.~\ref{FIDM_sol}.
In particular DM masses above $1\gev $ requires $ \lx < 10^{-13} $. 

If the scalar masses are $ \sim \mev $ and $\lx > 0$ then (\ref{mphi_bound})
requires $ \lx \ll 1 $, for example 
$\lx < 1.65 \times 10^{-7}$ and $\lx < 1.65 \times 10^{-5}$
for $\mvp= 100\mev$  and $\mvp= 1\gev $ respectively. Note also that from Fig.~\ref{cdm_sol_130} 
we conclude that there is no CDM solutions 
for $ \lx> 0 $ and $ \mvp < 70 \gev $ ; for smaller masses,
$ \lx $ is also very small and the situation reverts to the FIDM scenario.
One should however remember that solutions with $\lambda_x < 0$ are also 
possible; in this case Fig.~\ref{cdm_sol_130} implies there
are no CDM solutions for $\mvp \lsim 2 \gev $, then again
the only solution is the FIDM shown in Fig.~\ref{FIDM_sol}.

\section{Direct detection}
\label{Sec:dir-det}

In this section we  discuss constraints imposed on the model by searches
 for direct signals of DM particles scattering off nuclei. Even though 
the results of many experiments are available, here we will concentrate
on the constraints obtained by the XENON100 experiment \cite{Aprile:2011hi} as 
they impose strongest limits on DM - nucleon scattering 
cross-section $\sigma_{\rm DM-N}$ 
in the mass range of our interest;
at the end of this section we also comment on the recent results
for the CRESST-II experiment \cite{Angloher:2011uu}.

\begin{figure}[h]
\center
\includegraphics[height = 3 cm, width = 3 cm]{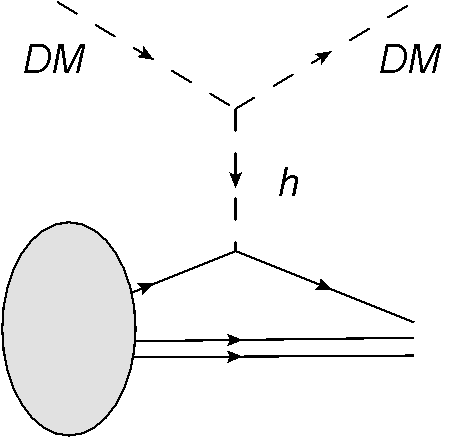}
  \caption{The Feynman diagram for the elastic scattering of $\vp$ off a nucleon.}
\label{dir_det}
\end{figure}

The relevant scattering amplitude is described by the Feynman diagram 
 in Fig.~\ref{dir_det}; the corresponding cross section is 
\beq
\sigma_{\rm DM-N} = \frac{1}{\pi} \frac{\lambda_x^2 m_n^2 \left( \sum_{q} f^{N}_{q} \right)^2 }{m_h^4 m_{\varphi}^2}
\label{dir_xs}
\eeq
where the sum runs over all quark flavors $q$, $m_n$ is the nucleon mass and $f^{N}_{q}$ are the nucleon form factors as defined in~\cite{Belanger:2008sj}.

\begin{figure}[h]
\includegraphics[height = 5 cm]{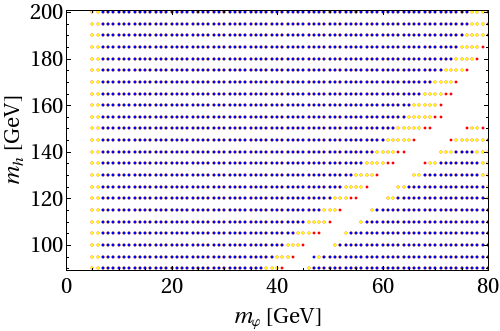}
\includegraphics[height = 5 cm]{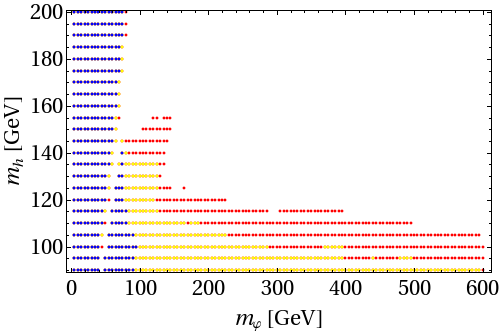}
  \caption{Constraint coming from the Xenon100 direct detection experiment combined with dark matter abundance in the case of CDM. Region forbidden for $N = 1$ is marked blue, for $N = 6$ blue and yellow, and $N=12$ yellow, blue and red. The left panel magnifies the region of low scalar mass $\mvp<80\gev$, while the right one extends the mass range up to $600\gev$.}
\label{direct_xenon}
\end{figure}

In Fig.~\ref{direct_xenon} we show the regions in the $(\mvp,\mh)$ plane
allowed by the XENON100 limits~\cite{Aprile:2011hi} on $\sigma_{\rm DM-N}$. In the left panel the low 
$\mvp$ region is magnified to illustrate the two allowed bands.  The one  corresponding to the resonance region $\mh \sim 2 \mvp$ (see also Fig.~\ref{cdm_sol_130}); for these values the annihilation is amplified to such an extent that (\ref{dm}) requires a  very small coupling $\lx$ so that this region is allowed
by the direct detection. Since the XENON100 data is available for $\mvp\ge 5\gev$ therefore the vertical band of masses below $5\gev$ is also allowed. The right panel shows the large allowed area available for increasing $m_{DM}$ where the sensitivity of XENON100 is reduced.

Recently a positive result for a direct detection in elastic scattering of DM particles off nucleons was 
announced~\cite{Angloher:2011uu} by the CRESST-II collaboration; with two points selected by a
maximum likelihood fit (Tab.~4 and Fig.~8 in \cite{Angloher:2011uu})
$M1:\; \{m_{\rm DM}=25.3\gev,~\sigma_{\rm DM-N}=11.6 \cdot 10^{-6}\pb\} $ and
$M2:\; \{m_{\rm DM}=11.6\gev,~\sigma_{\rm DM-N}= 3.7 \cdot 10^{-5}\pb \}$.
These cross sections are far above lower limit from XENON100 
experiment~\cite{Aprile:2011hi}, an issue that is yet to be
resolved. As a complement to the implications derived above using the XENON100
limits, we will also discuss the consequences the CRESST-II results
would have in constraining our model. 

\begin{figure}[h!]
\includegraphics[width = 7.8 cm]{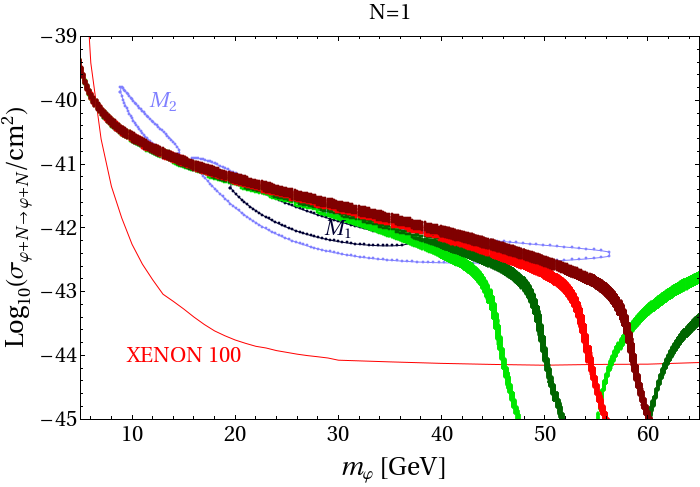}
\includegraphics[width = 7.8 cm]{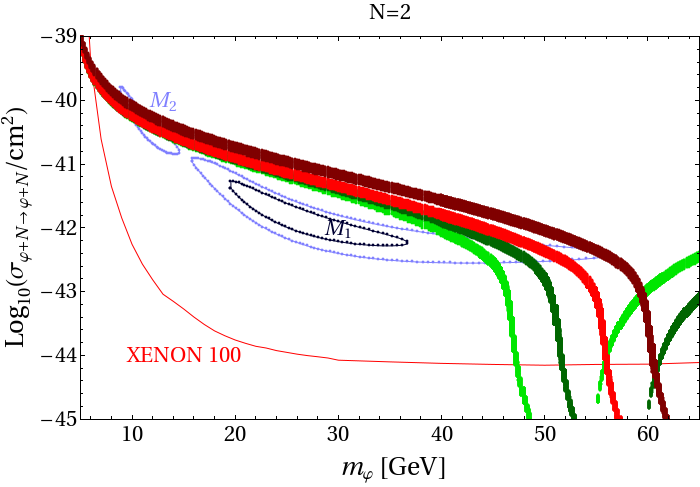}\\
\includegraphics[width = 7.8 cm]{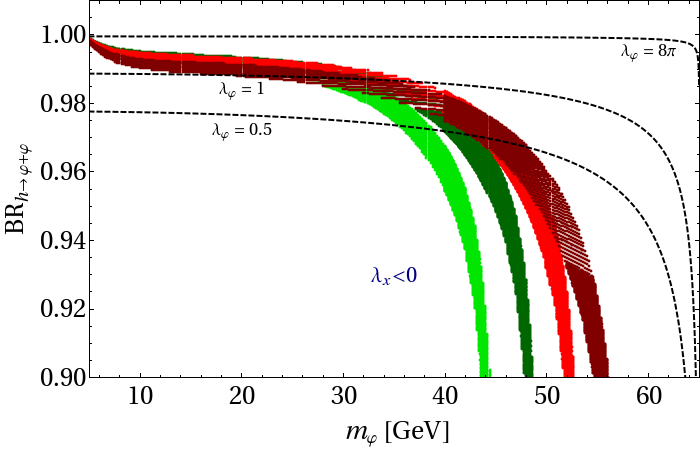}
\includegraphics[width = 7.8 cm]{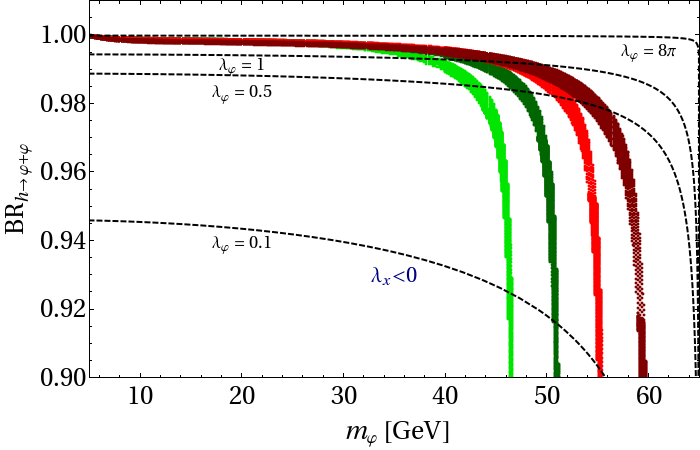}\\
\includegraphics[width = 7.8 cm]{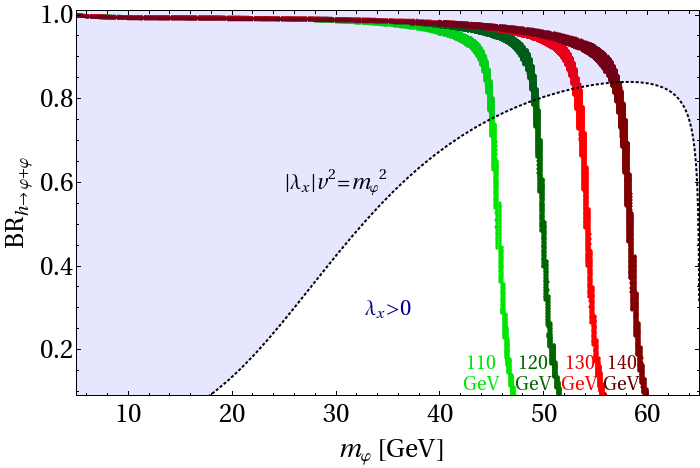}
\includegraphics[width = 7.8 cm]{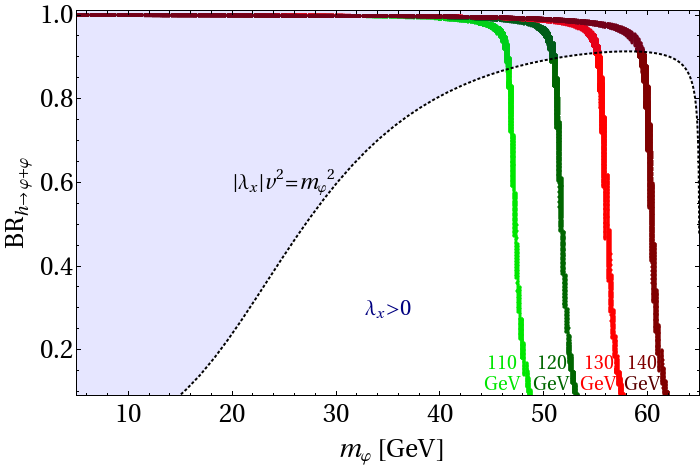}
\caption{Top panels: singlet-nucleon cross section for $N=1,2$ as a 
function of $ \mvp $ when (\ref{dm}) is obeyed, the 
color bands correspond $\mh = 110\gev$ (green), $120\gev$ (dark green), $130\gev$
(red), $140\gev$ (dark red); also shown the regions M1 and M2 favored
by the CRESST-II experiment \cite{Angloher:2011uu} at 1 and 2 $ \sigma $
level (black and blue closed curves, respectively), and the XENON100
limit~\cite{Aprile:2011hi}.
Middle panels: the branching ratio $BR_{h \to \vp \vp}$ for $N=1,2$
as a function of  $ \mvp $, 
when $ \lx < 0 $, for
four values of $ \mh $ (color coded as above); the
dashed lines show the value of $BR_{h \to \vp \vp}$
when $\lx$  saturates the 
stability bound (\ref{stab_con}) when the quartic $\vp$ self
coupling equals $\lvp=8\pi, 1, 0.5$ and $0.1$
and for $\mh=130\gev$: regions consistent with both (\ref{dm})
and (\ref{stab_con}) lies on the colored bands and below the
dashed curves.
Lower panels: the branching ratio $BR_{h \to \vp \vp}$ for $N=1,2$
as a function of  $ \mvp $, 
when $ \lx > 0 $, and for
four values of $ \mh $ (color coded as above); the dotted line
show the value of $BR_{h \to \vp \vp}$
when $\lx$  saturates the 
consistency bound (\ref{mphi_bound}): regions consistent
with both (\ref{dm}) and  (\ref{mphi_bound}) lie on the 
colored bands below the dotted line. }
\label{cresst_1}
\end{figure}

\begin{figure}[h!]
\includegraphics[width = 7.8 cm]{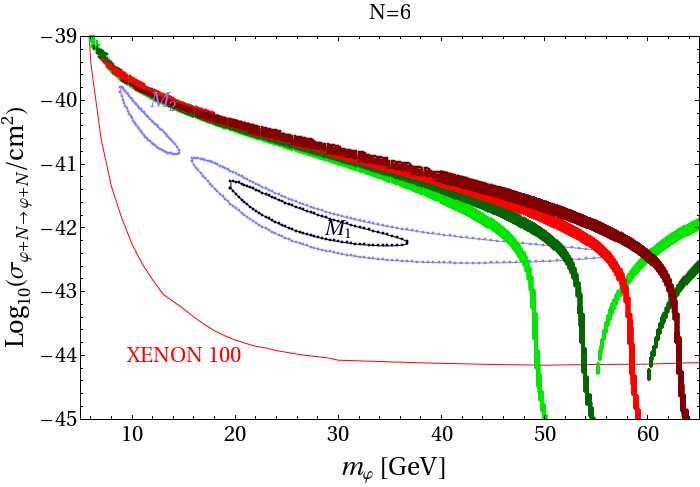}
\includegraphics[width = 7.8 cm]{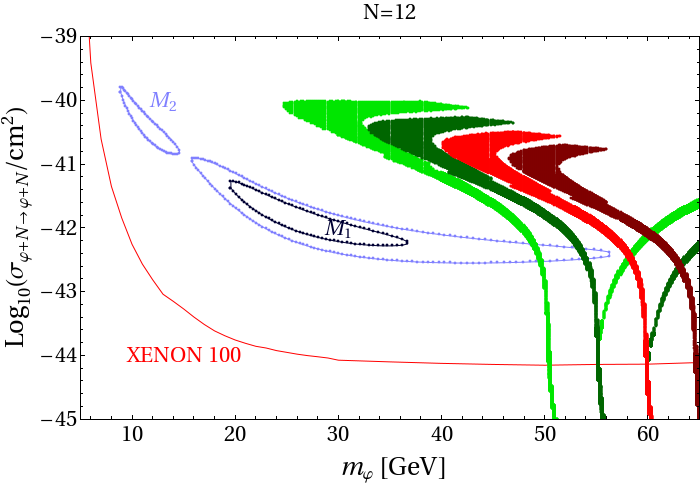}\\
\includegraphics[width = 7.8 cm]{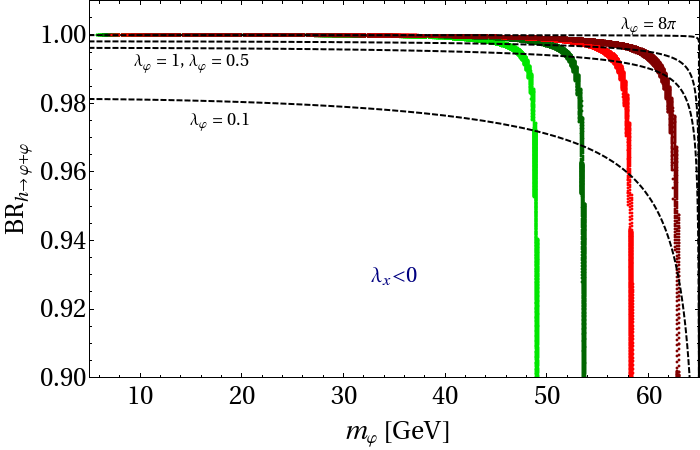}
\includegraphics[width = 7.8 cm]{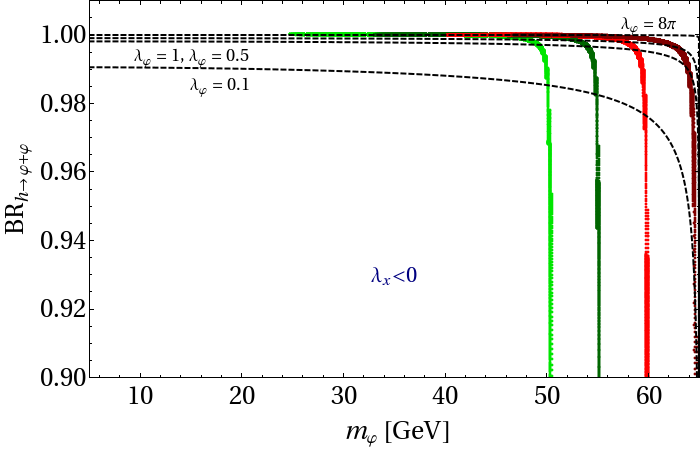}\\
\includegraphics[width = 7.8 cm]{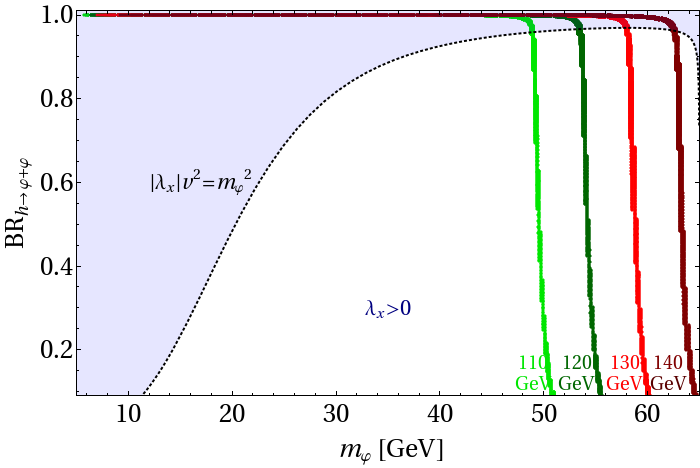}
\includegraphics[width = 7.8 cm]{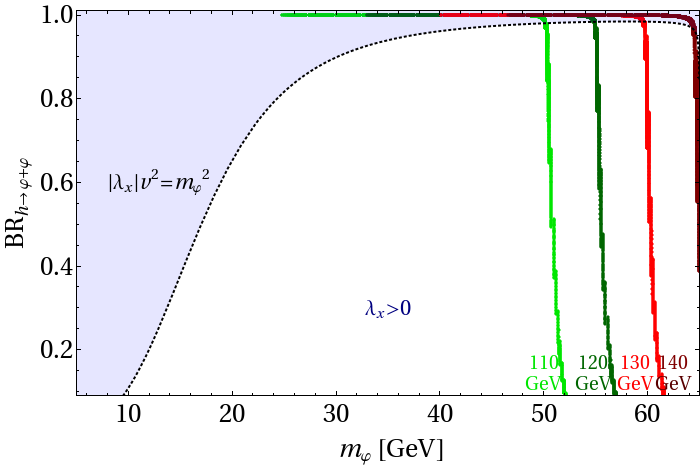}
\caption{Similar as in Fig.~\ref{cresst_1}, for $N=6$ and $N=12$.}
\label{cresst_2}
\end{figure}

For each choice of $ \mh $ and $N$ 
the requirement (\ref{dm}) determines a narrow allowed
band in the $ (\lx,\mvp) $ plane; an example is provided in
Fig.~\ref{cdm_sol_130} for the CDM case with
$\mh=130\gev$ and $N=1,6,12$. It is natural to
ask weather there exist $\mh$ (in the region allowed by the
present data) and $N$ such that the corresponding band 
contains the most likely points $M1$ and $M2$ found by
CRESST-II. The answer 
is contained in the top panels of Fig.~\ref{cresst_1}, where the
regions preferred by
CRESST-II (within $2 \sigma $) are superimposed with predictions 
of our model for $N=1,~2$ and $\mh=110-140\gev$
when satisfying (\ref{dm}) at the $ 3 \sigma$ level.
As it is seen from the figure both regions are consistent with the model.
For larger values of $N$ our model is consistent only with the 2$\sigma$ $M1$ contour for $ \mvp \sim 50 \gev $. If confirmed, these data would provide a strong restriction on the scenario discussed in this paper; in particular a tightening of the allowed regions around the current central values would require $ N \lsim 5 $.
Moreover it has immediate and dramatic
consequences for Higgs boson searches at Tevatron or LHC since
the
invisible $h\to \vp \vp$ decay has a substantial 
branching ratio for the parameter regions
consistent with $M1$ or $M2$.
This is illustrated in the middle and bottom panels of
Fig.~\ref{cresst_1}, where the branching ratio is
plotted against scalar mass $\mvp$. 

When $ \lx < 0 $,
the constraint (\ref{mphi_bound}) generates no
restrictions and the branching ratio can be very large,
$BR_{h \to \vp\vp}\sim 1$ (see middle panels of Figs.~\ref{cresst_1} and \ref{cresst_2}). 
In these figures we also plot 
maximum value of $BR_{h \to \vp \vp}$
calculated for $|\lx|$ that saturates the vacuum
stability bound (\ref{stab_con}) on $|\lx|$ for a given quartic
$\vp$ self coupling $\lvp=8\pi, 1, 0.5$ and $0.1$,
when $\mh=130\gev$; the regions above corresponding
dashed curves are excluded by this constraint.

For $ \lx > 0 $ 
the constraint (\ref{mphi_bound}) is important and
excludes the region above  the dotted curve shown in the 
bottom panels of Fig.~\ref{cresst_1} and \ref{cresst_2}. Within the 
allowed region 
the $BR_{h \to \vp\vp}$ is smaller, but it can still
reach $\sim 0.8 $.

Note, that if one anticipates the cutoff $\Lambda$ to be below
$\sim 10^4\gev$ (as we do in this work)
then the triviality upper limit for low $N$ is roughly $\lvp \lsim
5$, so the branching ratio could easily reach
even $99\%$ without any conflict with triviality. 
It is also worth mentioning here that if $\lx > 0$, then 
{\underline small value of $\mvp$ in the $M1$ and $M2$ regions together with
the consistency condition (\ref{mphi_bound}) imply
$0.06 < \lx < 0.24$, which is too
small to ameliorate the SM fine tuning problem (see
Sec.~\ref{Sec:fin-tun})}. On the other hand, if $\lx < 0$,
the vacuum stability bound (\ref{stab_con}) implies
$|\lx| < 0.85$ for
$\lvp \le 8 \pi$ and $110 \gev < \mh < 141\gev$; then, as seen from 
Fig.~\ref{FT_neglx}, the allowed values for the
cutoff $ \Lambda $ are smaller than for the SM.
Therefore $\lx < 0$ does not help solving the hierarchy problem regardless of
the cosmological constraints.

From the top panels of Figs.~\ref{cresst_1} and \ref{cresst_2} one observes that
for $1 \leq N \leq 12$ low Higgs boson masses ($\mh=110-120\gev$)
are preferred by the $M1$ region of heavier $\vp$ (larger $N$ favors smaller $\mh$). 
For the region of lighter $\mvp$ ($M2$) roughly all Higgs boson
masses fit data equally well, however only for $N=1$ and 2.

In conclusion, the CRESST-II positive result, if confirmed, would
imply (within the
singlet extension of the SM) that the Higgs boson could decay mainly
invisibly, escaping discovery at both Tevatron and LHC. The option of
invisibly decaying Higgs boson
have been considered previously in \cite{Binoth:1996au}.

\section{Self-interacting DM}
\label{Sec:self-int}

In spite of its many successes, the standard CDM cosmological model
is also 
facing some difficulties when compared with recent observations.
For instance,
high-resolution N-body simulations have shown that the model
generates a  cusps in the DM density distribution in
central regions of galaxies~\cite{Navarro:1996gj}.
Another discrepancy concerns the number of subhalos predicted by
the model, which is
at least factor of ten larger than the
observed~\cite{Klypin:1999uc} number. Self-interacting DM was
proposed by Spergel
\&\ Steinhardt~\cite{Spergel:1999mh} to cure these problems.  
Within this model, dark matter particles experience weak,
non-dissipative collisions on scales of kpc to
Mpc for typical galactic densities. This effect then
generates a soft core in the inner regions of the dark
halos, and it also ameliorates the overabundance of subhalos. 
The key requirement is that the mean free
path of DM particles should be between 1 kpc and
1 Mpc in regions where the dark matter density is about
$0.4\gev/{\rm cm}^3$. This model has attracted much
attention~\cite{Wandelt:2000ad}.

\begin{figure}[t!]
\center
\includegraphics[height = 3 cm, width = 3 cm]{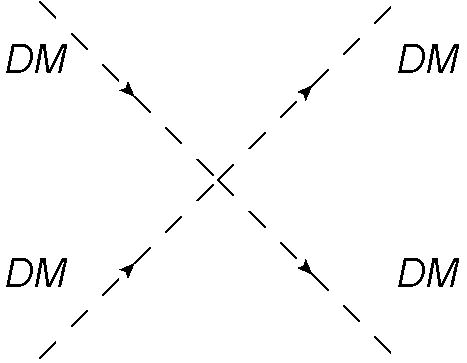}\hspace*{1cm}
\includegraphics[height = 3 cm, width = 3 cm]{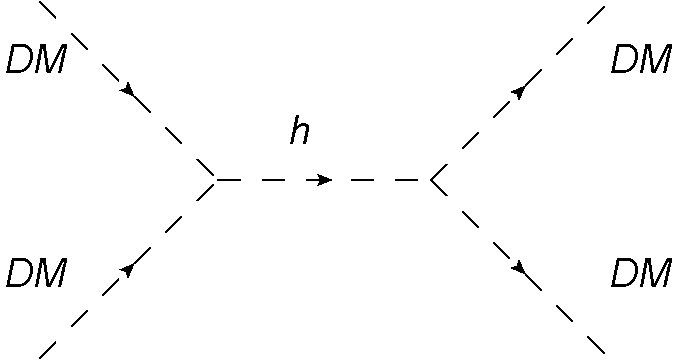}\hspace*{1cm}
\includegraphics[height = 3 cm, width = 3 cm]{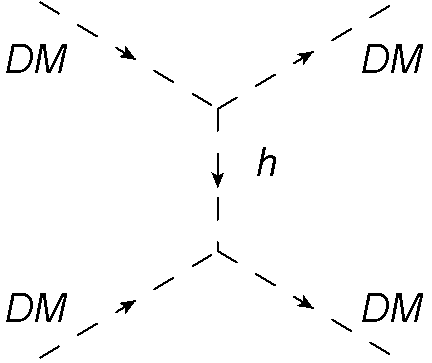}
  \caption{Feynman diagrams contributing to the elastic $\vp\vp$ scattering.}
\label{dm-elas-scatt}
\end{figure}

In order to discuss this idea quantitatively we define
the elastic scattering cross section per unit mass
for DM consisting of $N$ species. We then imagine a clump of DM particles
scattering off another such clump:
\beq
\frac{\sigma_{DM}}{m_{DM}}=\frac{\sum_{i=1}^{N} n_{i}  \sum_{j,k,l}^{N} \sigma_{i j \rightarrow k l}}{\sum_{i=1}^{N} n_{i} m_{i}}
\eeq
where $i,j, $ etc. label the DM flavor (in our case, the components of 
$ \vp $), $m_i $ denote the corresponding masses, 
$n_{i}$ the number density of $i$-th particles in the clump,
and $  \sigma_{i j \rightarrow k l} $ the 2 on 2 cross sections.
Because of the $O(N)$ symmetry  the $n_i =n $ and $m_i = \mvp$
are flavor-independent, therefore
\beq
\frac{\sigma_{DM}}{m_{DM}}=\frac{ N \sum_{j,k,l}^{N} \sigma_{1 j \rightarrow k l}}{ N \mvp} = 
\frac{  \sigma_{1 1 \rightarrow 11}+ (N-1)\sigma_{1 1 \rightarrow 22}+(N-1)\sigma_{12 \rightarrow 12} }{ m_{\varphi}}
\label{unit_sigma}
\eeq
We will define an auxiliary self-interaction cross section:
\beq
\sigma_{\vp \vp} (\lambda_1,\lambda_2,\lambda_3) =
\frac{\left(
\mh^4 \lambda_1 + 32 \mvp^2 v^2 \lambda_3^2 - 4\mh^2 ( \mvp^2 \lambda_1 + v^2 \lambda_2^2 + 2 v^2 \lambda_3^2)
\right)^2}{128 \pi  \mvp^2 \mh^4 (\mh^2 - 4\mvp^2)^2} 
\eeq
where $\lambda_1$ is the coupling of the direct quartic interaction, $\lambda_2$ is the s-channel Higgs exchange coupling and $\lambda_3$ corresponds to the t- and u-channel Higgs exchange. Now the relevant cross sections are: 
\begin{eqnarray}
\sigma_{1 1 \rightarrow 11} &=& \sigma_{\vp \vp } (\lvp, \lx,  \lx) \label{s1111}\\
\sigma_{1 1 \rightarrow 22} &=& \sigma_{\vp \vp } (\frac{\lvp}{3},  \lx, 0) \label{s1122}\\
\sigma_{12 \rightarrow 12} &=& 2 \, \sigma_{\vp \vp} (\frac{\lvp}{3}, 0,  \lx) 
\label{s1212}
\end{eqnarray}
where the factor $\frac{1}{3}$  in (\ref{s1212}) and (\ref{s1122}) comes from the combinatorics of the relevant diagrams and the factor of 2 in (\ref{s1212}) corresponds to the absence of two identical particles in the 
final state. The relevant diagrams  are shown in Fig.~\ref{dm-elas-scatt}.

In terms of the cross section per unit mass (\ref{unit_sigma}), 
the Spergel \&\ Steinhardt hypothesis requires that
\beq
2.05 \cdot 10^3 \gev^{-3} \lsim \frac{\sigma_{DM}}{m_{DM}} \lsim 2.57 \cdot 10^4 \gev^{-3}
\label{SS_con}
\eeq
In this section we will investigate whether this constraint is consistent with 
all the other restrictions that we  have imposed on the multi-singlet extension of the SM.

\begin{figure}[th!]
\center
\includegraphics[height = 4.9 cm, width = 4.9 cm]{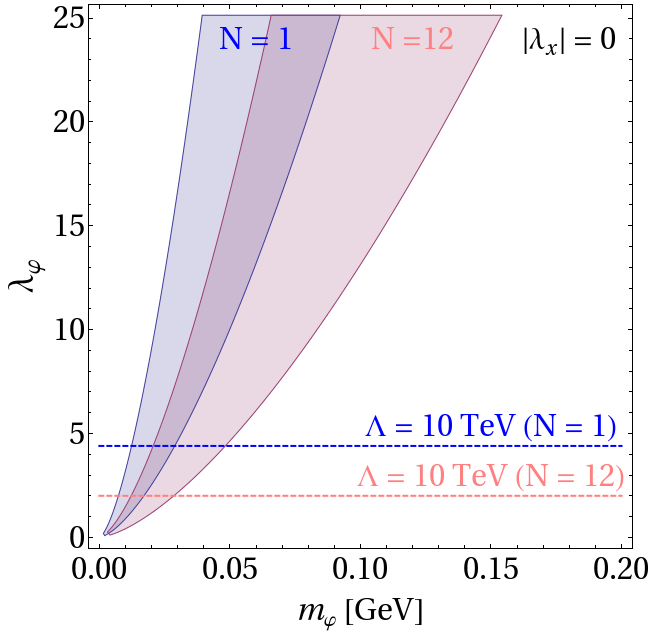}
\includegraphics[height = 4.9 cm, width = 4.9 cm]{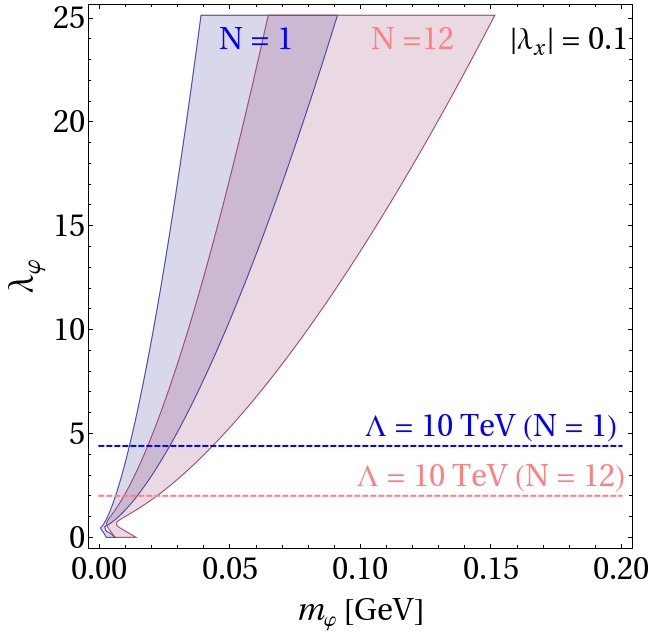}
\includegraphics[height = 4.9 cm, width = 4.9 cm]{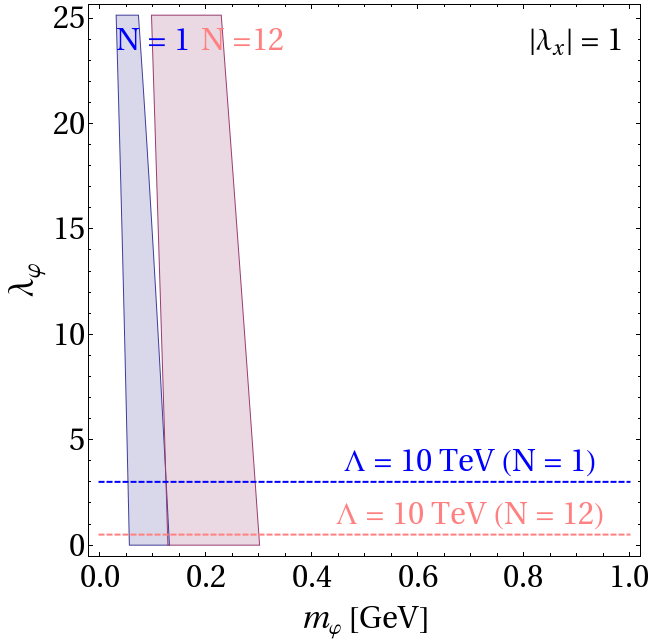}
  \caption{Regions in $(\lvp,\mvp)$ space allowed by the Steinhard-Spergel
  hypothesis for $N=1$ (blue/darker), $N=12$ (pink/brighter) region and the Higgs-boson mass 
  fixed at $\mh=130 \gev$. The left,
  middle, and right panels correspond to $\lx=0,\,0.1$ and $1.0$, respectively.
  The horizontal lines shows upper limits on $\lambda_{\varphi\, 0}$ allowed by the triviality
  condition for the cutoff $\Lambda=10\tev$ and $N=1,12$.}
\label{ss}
\end{figure}

For a given $N$ and $\mh$, the condition
(\ref{SS_con}) defines an allowed region in ($\lx, \lvp$, $\mvp$) space;
the projections onto the $(\mvp,\lvp)$ plane for fixed
 $ |\lx| = 0,\, 0.1,\, 1$ with
$N=1,\, 12$ and $\mh=130 \gev$, are shown in
Fig.~\ref{ss}.
In the plots $\lvp$ varies form $0$ up to $8\pi$,
the maximum value allowed by perturbative unitarity (\ref{unit_con}).
It is clear that the condition (\ref{SS_con}) poses
a very strong constraint on $\mvp$, which is restricted to small
values, generally well below $ 1 \gev $ for $ |\lx| \lsim 1 $ and 
$ N \lsim 10 $ ($ \mvp $ values $ \sim 100 \gev $ are allowed only for
for $ |\lx| \sim 10 $ and $ N \gsim 100 $); similar
results were obtained in other versions of scalar DM models
\cite{Bento:2001yk,McDonald:2001vt,Holz:2001cb}.

If, in addition, one requires that the scale of
physics beyond the singlet extension is above $10\tev$, then
Landau poles are forbidden below 
this value for $\Lambda$~\footnote{In our derivation of $ \sigma_{DM}$
we neglected radiative corrections: because
of the  small allowed values of $ \mvp $, we expect that the
effects from the renormalization group evolution of the couplings
will be small, and for this reason in this section we
do not differentiate between $ \lvp $ and its initial
value (\ref{initial3}).}. As a consequence, the limits on $\mvp$ are
much stronger, for example in the case of $\lx=0$ we obtain: $\mvp \lsim
0.025\gev ~(N=1)$ or $\mvp < 0.045\gev~ (N=12)$.

As observed from Fig.~\ref{cdm_sol_130} the values of $\mvp$ that are
required by the Spergel \& Steinhardt condition (\ref{SS_con}) are not
compatible with the CDM case. The only viable option is the
FIDM. In this case, from Fig.~\ref{FIDM_sol}
one can see that $\mvp \sim 0.01-0.05\gev$ corresponds to $\lx
\sim 10^{-10}-10^{-12}$ for $N=1-12$, and only the first panel 
in Fig.~\ref{ss} is consistent with the DM abundance.
Thus, within our model, the Spergel \& Steinhardt hypothesis 
is consistent only with very light DM particles ($\mvp \lsim 0.01-0.05\gev$)
that are very weakly coupled to the SM ($\lx\sim 10^{-12}-10^{-10}$).

\section{Summary and conclusions}
\label{Sec:sum}

We have considered an extension of the Standard Model 
that contains a set of $N$ real scalar
gauge singlets $\vp$ that transform
as the fundamental representation of a global $O(N)$ symmetry, under
which all SM particles are singlets. This global symmetry remains unbroken, so
that $\vp$ is a stable candidate for Dark Matter. The constraints on the model parameters implied
by tree-level vacuum stability, unitarity and triviality were discussed. 
We have also investigated to what extent the presence of extra scalars could ameliorate
the fine tuning of quadratic corrections to the Higgs boson mass. 

The restrictions imposed by abundance of DM were also presented;
in particular we showed that the recent CRESST-II data 
for DM-nucleus scattering, if confirmed, would imply that for this model
the standard Higgs boson decays predominantly into pairs of Dark Matter scalars.
It that case discovery of the Higgs boson at LHC and Tevatron is impossible.
For $N=1$ the most likely mass of the dark scalars lies in the range 
$15\gev \lsim \mvp \lsim 50\gev$ and $BR(h\to \vp\vp)\sim 96\%$. 
If $N \gsim 2$, the scalars have to be heavier: 
$50\gev \lsim \mvp \lsim 60\gev$, 
and $BR(h\to \vp\vp)\sim 98\%$. 
We have also shown that
the Spergel-Steinhardt solution of the Dark Matter density
cusp problem restricts the parameters
to a Feebly Interacting Dark Matter region with
$\lx \sim 10^{-12} - 10^{-10}$ and scalar masses in the range 
$\mvp \sim 0.01-0.05\gev$.

\vspace{.5in}
\acknowledgments

This work has been partially financed by the National Science Centre (Poland)
as a research project, decision no DEC-2011/01/B/ST2/00438; and 
by the U. S. Department of Energy under Grant No. DEAC02-06CH11357.
The authors thank both the XENON and CRESST-II collaborations for providing 
relevant data concerning the allowed regions in the 
$(\sigma_{DM\,N\to DM\, N},m_{DM})$ space
that were adopted for plots shown in this work.
AD acknowledges financial support from the project "International PhD
Studies in Fundamental Problems of  Quantum Gravity and Quantum Field
Theory" of Foundation for Polish Science, co-financed from the program IE
OP 2007-2013 within European Regional Development Fund.


\end{document}